\documentclass[aip,jcp,reprint]{revtex4-1}
\usepackage{graphicx}
\usepackage{bm}

\begin{document}
\newcommand{\X}{\mathbf{x}}
\newcommand{\Y}{\mathbf{y}}
\newcommand{\Z}{\mathbf{z}}
\newcommand{\F}{\mathbf{f}}
\newcommand{\Data}{\{y_i^{\OBS}\}}
\newcommand{\DATA}{\{Y_i^{\OBS}\}}
\newcommand{\OBS}{(\mathrm{obs})}
\newcommand{\YiEns}{y}
\newcommand{\Popt}{p^{\mathrm{(opt})\!}}
\newcommand{\Wopt}{w_\alpha^{\mathrm{(opt})\!}}
\newcommand{\WoptG}{w_\gamma^{\mathrm{(opt})\!}}

\title{Bayesian ensemble refinement by replica simulations and reweighting}

\author{Gerhard Hummer}
\email[Author to whom correspondence should be addressed. Electronic
mail: ]{gerhard.hummer@biophys.mpg.de}

\author{J\"urgen K\"ofinger}
\email[Submitted to The Journal of Chemical Physics]{}

\affiliation{$^1$Department of Theoretical Biophysics, Max Planck Institute of
  Biophysics, Max-von-Laue Str. 3, 60438 Frankfurt am Main, Germany}

\begin{abstract}
  We describe different Bayesian ensemble refinement methods, examine
  their interrelation, and discuss their practical application.  With
  ensemble refinement, the properties of dynamic and partially
  disordered (bio)molecular structures can be characterized by
  integrating a wide range of experimental data, including
  measurements of ensemble-averaged observables.  We start from a
  Bayesian formulation in which the posterior is a functional that
  ranks different configuration space distributions.  By maximizing
  this posterior, we derive an optimal Bayesian ensemble
  distribution. For discrete configurations, this optimal distribution
  is identical to that obtained by the maximum entropy ``ensemble
  refinement of SAXS'' (EROS) formulation.  Bayesian replica ensemble
  refinement enhances the sampling of relevant configurations by
  imposing restraints on averages of observables in coupled replica
  molecular dynamics simulations.  We show that the strength of the
  restraint should scale linearly with the number of replicas to ensure
  convergence to the optimal Bayesian result in the limit of infinitely many
  replicas.  In the
  ``Bayesian inference of ensembles'' (BioEn) method, we combine the
  replica and EROS approaches to accelerate the convergence.  An
  adaptive algorithm can be used to sample directly from the optimal
  ensemble, without replicas. We discuss the incorporation of
  single-molecule measurements and dynamic observables such as
  relaxation parameters.  The theoretical analysis of different
  Bayesian ensemble refinement approaches provides a basis for
  practical applications and a starting point for further
  investigations.
\end{abstract}

\maketitle

\section{Introduction}

The problem of ensemble refinement \cite{Boomsma:14:1} becomes
increasingly important as structural biology enters a new era in which
dynamic and partially disordered biomolecular structures come into
focus.\cite{wwPDB:Structure:2015,Boura:11,Sali:Science:13} Such
systems play central roles in biology, both in functional cellular
processes ranging from signal transduction to the formation of large
cellular structures, and in disease, including neurodegenerative
diseases such as Parkinson's and Alzheimer's.  A broad range of
methods have been developed to refine models of (bio)molecular
structures against experimental data from X-ray crystallography,
nuclear magnetic resonance (NMR) spectroscopy, electron microscopy
(EM), solution X-ray or neutron scattering (SAXS, SANS), and other
methods.  By and large, these refinement methods operate under the
assumption that a single or a few well ordered structures should
account for all the measurements.  However, refinement of a single (or
possibly a few) copies is not appropriate in systems with significant
disorder.  For unfolded \cite{Camilloni:14} or intrinsically
disordered proteins
(IDP),\cite{Fisher:10,Fisher:11,Sanchez-Martinez:14} such as the
$\alpha$-synuclein peptide involved in Parkinson's
disease,\cite{Dedmon:05,Mantsyzov:14,Mantsyzov:15} we expect that a
very broad range of structures is present in solution.  None of these
structures may individually satisfy all measurements, and even if one
did, it may be highly atypical.  Instead, most observables accessible
to experiment report on averages over the entire ensemble of
structures, and as such only the appropriate average over a model
ensemble should match the experiment.

\begin{figure}[htb]
  \centering
  \includegraphics[width=8cm]{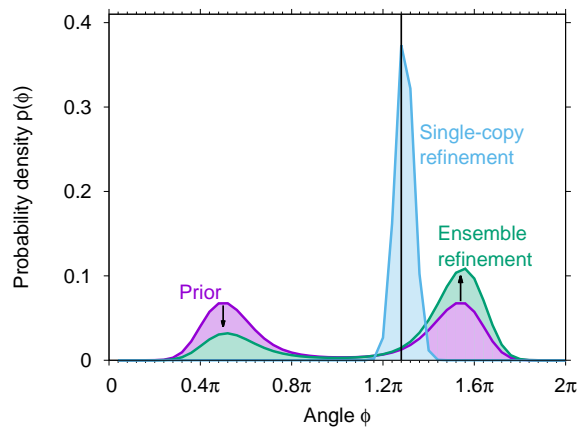}
  \caption{Illustrative comparison of refined probability densities
    $p(\phi)$ of a dihedral angle $\phi$
    from single-copy refinement [blue;
    Eqs.~(\ref{eq:1}-\ref{eq:3}); every member of the ensemble is expected to
satisfy the measurement individually] and ensemble refinement [green;
    Eq.~(\ref{eq:34}); the ensemble
average is expected to satisfy the measurement].
   The prior or reference distribution (magenta) used
   in the refinements is bimodal,
   i.e., with two dominant rotamer states.
    As indicated by the vertical black line, the
    observable is $y(\phi)\equiv \phi=1.28\pi$ for single-copy refinement and
    $Y\equiv\overline{\phi}=1.28\pi$ for ensemble refinement, with
    ``experimental'' error $\sigma=0.04\pi$ in both cases, and
    $\theta=1$. Arrows indicate the changes in the relative weights
    of the two rotamers in
   the optimal Bayesian ensemble refined distribution.
   \label{fig:1}}
\end{figure}

Ensemble refinement is a challenging inverse problem in which one aims
to characterize the high-dimensional configuration space of a
molecular system on the basis of limited experimental information. It
is therefore essential that ensemble refinement methods can properly
integrate data from a broad range of experiments
\cite{Sali:Science:13,Schroder:15,wwPDB:Structure:2015} that may
report on molecular size and shape (e.g., from SAXS, SANS, or
hydrodynamic measurements \cite{Rozycki:11,Boura:11,Francis:11}), the
proximity (e.g., from cross-links) or distance between atoms and
residues [e.g., from fluorescence resonant energy transfer (FRET),
NMR,\cite{Scheek:91,Lange:08,Olsson:14} including nuclear Overhauser
effects (NOE), or double electron-electron resonance (DEER)
measurements \cite{Boura:11,Boura:12}], the local chemical environment
and structure (e.g., from NMR chemical shifts \cite{Francis:11} and
J-couplings \cite{Mantsyzov:14,Mantsyzov:15} or X-ray absorption
spectroscopy), all the way to measures of the global structure (e.g.,
from X-ray crystal diffraction or electron microscopy
\cite{Cossio:13,Sali:Science:13,Schroder:15}).  Taking into account
the uncertainties of the different experiments
\cite{Schneidman-Duhovny:14} is critical for the construction of a
properly weighted configurational ensemble.

Inverse problems are typically ill-conditioned, i.e., sensitive to
input parameter variations, and underdetermined. Such problems with
high sensitivity and low data-to-parameter ratios are usually
tackled through regularization, for instance by assuming near-uniform
and smooth solutions. Bayesian statistics offers a particularly
elegant route for the inference of probabilistic models from data
(see, e.g., Ref. \onlinecite{McKayBook} for a general overview and
Ref. \onlinecite{Rieping:05} for a pioneering application to biomolecular
studies).  In effect, the assumed prior distributions of the model
parameters serve as regularizing factors,
\begin{equation}
  \label{eq:0}
  p(\textrm{model}|\textrm{data})\propto p(\textrm{data}|\textrm{model}) p_0(\textrm{model}),
\end{equation}
written as a proportionality without the normalizing factor.
$p(\textrm{model}|\textrm{data})$ is the posterior distribution of the
model, and $p_0(\textrm{model})$ is the prior that expresses our
expectations on the model and its parameters in the absence of new data.
$p(\textrm{data}|\textrm{model})$ is the conditional probability of
observing the data given the model, which for given data is the
likelihood of the model.  Consequently, we will in the following refer
to $p(\textrm{data}|\textrm{model})$ as the likelihood function.
Importantly, in the absence of new data (or for non-informative data), one
simply recovers the prior.

The importance of ensemble refinement is best illustrated by a simple
example that anticipates some of the theoretical developments in this
work. Figure~\ref{fig:1} contrasts the stark differences in the
results for single-copy and ensemble refinements of a simple model
system with a prior or reference distribution with two dominant rotamers.
In single-copy refinement, we determine how
well each dihedral angle $\phi$ individually agrees with the observation
$y^{\OBS}=\phi^{\OBS}=1.28\pi$. This posterior probability $p(\phi|\mathrm{data})
\equiv p(\phi|y^{\OBS})$ is
concentrated in a sharp peak around the target value. By contrast, in
ensemble refinement we seek a probability density $p(\phi)$ that is
consistent with the observed average $Y^{\OBS}\equiv\overline{\phi}$.  This
$p(\phi)\equiv\Popt(\phi)$ retains the character of the reference
distribution that reflects the underlying physics, while redistributing
some population from one rotamer to the other.

Here, we will describe both formal and practical approaches toward
inferring ensemble distributions from diverse data.  We will formulate
the ensemble refinement problem first formally in a Bayesian framework
in which the posterior is a functional that quantifies the relative
probability of different ensemble probability densities $p(\X)$ for
configurations $\X$.  Experimental uncertainties
\cite{Schneidman-Duhovny:14} are taken into account from the outset,
which allows us to combine data from a variety of measurements.  We
then study algorithms to realize Bayesian ensemble refinement in
practice.  First, we will describe a method with which existing
ensembles can be reweighted to match experiment.  By variational
maximization of the Bayesian posterior functional over the ensemble
probability densities $p(\X)$, we will derive an optimal Bayesian
ensemble density $\Popt(\X)$, Eq.~(\ref{eq:34}), for the continuous
case and Eq.~(\ref{eq:18}) for the discrete case.  Applied to
sub-ensembles drawn according to the prior, this reweighting method
turns out to be equivalent to the maximum entropy refinement procedure
in the ensemble refinement of SAXS (EROS) method \cite{Rozycki:11}
(which is different from the ``ensemble refinement with orientational
restraints'' method with the same acronym \cite{Lange:08}).  In the
limit of infinite sample size, the reweighting method converges to the
optimal Bayesian ensemble refinement.  Then, we will describe a
Bayesian replica ensemble refinement method to perform ensemble
refinement on the fly by running molecular simulations of identical
copies of the system with a bias on the averages calculated over these
replicas.  If the biasing potential is proportional to chi-squared (as twice
the negative log-likelihood for Gaussian errors) scaled by the number
of replicas $N$ [see Eq.~(\ref{eq:21})], one recovers the optimal
Bayesian ensemble refinement [Eq.~(\ref{eq:34})] in the limit of an
infinite number of replicas. At the other extreme, in the limit of a
single replica, the common-property refinement [Eq.~(\ref{eq:7})] is
recovered, in which every member of the ensemble is expected to
satisfy the measurements individually, not just in the ensemble
average.  To speed up the convergence to the optimal Bayesian
distribution with increasing number of replicas $N$, we show how EROS
and replica refinement (as well as other ensemble-biased simulation
methods) can be combined with the help of free-energy reweighting
methods, resulting in the ``Bayesian inference of ensembles'' (BioEn)
method.  An adaptive algorithm designed to sample directly from the
optimal ensemble distribution, without multiple replicas, is presented
in an Appendix.

To illustrate the formal theory and the practical replica simulation
approaches, we will introduce analytically or numerically tractable
models of ensemble refinement. The solutions obtained for these models
allow us to assess the mutual consistency of the methods, and to
demonstrate the need for a size-consistent treatment in the Bayesian
replica ensemble refinement with respect to the number of replicas.
We also sketch how dynamic properties can be integrated in ensemble
refinement, albeit approximately, and how the parameter expressing the
confidence in the reference ensemble distribution can be chosen.  We
conclude by a summary of the main results and a discussion of possible
applications, including the optimization of potential energy functions
used for molecular simulations.

\section{Theory}

\subsection{Bayesian single-copy refinement in configuration space}

Before venturing into ensemble refinement, we introduce notation and
the general framework in the context of the more familiar single-copy
refinement.  Here one assumes that a single configuration can explain
all measured data.  Different configurations can then be ranked, in a
probabilistic manner, by their respective abilities to do so.

In the following, we will use $\X$ to denote individual
configurations.  In a typical application to a molecular system, $\X$
could be the $3n$-dimensional vector
$\X=\left\{\mathbf{r}_1,\mathbf{r}_2,\ldots,\mathbf{r}_n\right\}$ of
the Cartesian coordinates $\mathbf{r}_i$ of the $n$ atoms.  In
single-copy refinement, we assume that one would ideally (i.e.,
without error) measure values $y_i(\X)$ of observable $i$, with
$i=1,2,\ldots,M$, for a given configuration $\X$.  The actual values
observed (measured) are $y_i^{\OBS}$.  By contrast, in ensemble
refinements described below, the measured values of the observables
will instead depend on the distribution over the entire configuration
space, not just on a single configuration $\X$.

Using a reference distribution $p_0(\X)$ as a prior, in single-copy
refinement we want to construct a posterior $p(\X|\mathrm{data})$ in
configuration space that ranks configurations $\X$ by their
consistency with both experimental measurements and prior.  The prior
$p_0(\X)$ could for instance be the Boltzmann distribution for a
simulation model described by a particular potential energy function
$U(\X)$, i.e., $p_0(\X)=\exp[-\beta U(\X)]/\int d\X' \exp[-\beta
U(\X')]$ at reciprocal temperature $\beta=1/k_BT$ with $k_B$ the
Boltzmann constant and $T$ the absolute temperature, or a statistical
distribution of conformers of the Protein Data
Bank.\cite{Mantsyzov:14,Mantsyzov:15} The normalized posterior
distribution according to Eq.~(\ref{eq:0}) is then
\begin{equation}
  \label{eq:1}
  p(\X|\Data) = \frac{p_0(\X) p(\Data|\X)} {\int d\X' p_0(\X') p(\Data|\X')},
\end{equation}
where $p(\Data|\X)$ is the likelihood of $\X$ for data given as a set
of $M$ measured values,
$\Data\equiv\{y_1^{\OBS},y_2^{\OBS},\ldots,y_M^{\OBS}\}$.  The
posterior $p(\X|\Data)$ gives the probability density that
configuration $\X$ is the single configuration underlying the data.

In cases where the statistical errors are Gaussian, we define the likelihood
function is
\begin{equation}
  \label{eq:2}
  p(\Data|\X) \equiv e^{-\chi^2(\X)/2},
\end{equation}
where
\begin{equation}
  \label{eq:3}
  \chi^2(\X) = \sum_{i=1}^M \frac{\left[y_i(\X)-y_i^{\OBS}\right]^2} {\sigma_i^2}.
\end{equation}
and $\sigma_i$ is the standard deviation of measurement $i$. For
simplicity, we assume in Eq.~(\ref{eq:3}) that the errors in the
different measurements $i$ are uncorrelated. In the more general case
of correlated errors, one can use
\begin{equation}
  \label{eq:4}
  \chi^2(\X) = \mathbf{\delta y}^T(\X) \bm{\Sigma}^{-1} \mathbf{\delta y}(\X)
\end{equation}
where $\mathbf{\delta y}$ is a vector of deviations, with elements
$\delta y_i(\X) = y_i(\X) - y_i^{\OBS}$, and $\bm{\Sigma}$ is the
symmetric covariance matrix of the statistical errors (where for uncorrelated
errors $\Sigma_{ii}=\sigma_i^2$ and $\Sigma_{ij}=0$ for $i\ne j$).  Note that the
measurements $i$ can be from different measurements (say, NMR and
single-molecule FRET) or from the same measurement (say, intensities
at different wave vectors in a SAXS measurement).

In practice, single-copy Bayesian refinement can then be performed by
sampling directly from the posterior $p(\X|\mathrm{data})$, e.g., by
running equilibrium simulations with an effective energy function
$U_\mathrm{eff}(\X)=U(\X)-\beta^{-1}\ln p(\mathrm{data}|\X)$.
Alternatively, representative configurations can first be sampled
from the reference distribution $p_0(\X)$ and then reweighted by the
likelihood according to Eq.~(\ref{eq:1}).

\subsection{Bayesian ensemble refinement in probability density space}

In an alternative Bayesian formulation, we think of $p(\X)$ not as a
posterior $p(\X|\mathrm{data})$ ranking individual configurations $\X$
with respect to their mutual consistency with prior and data, but as
an actual probability density of $\X$ in configuration space defining
an ensemble.  As a consequence, prior, likelihood, and posterior
become functionals of the probability density $p(\X)$ in configuration
space.  We note that such ``hyperensembles'' have been studied by
Crooks as models of nonequilibrium states.\cite{Crooks:2007}
Functional approaches are also used in variational Bayesian methods.\cite{McKayBook}

To construct a prior in the space of probability densities $p(\X)$,
with $p(\X)>0$ and $\int d\X\,p(\X)=1$, we use the Kullback-Leibler
divergence or relative entropy with $p_0(\X)$ as reference
distribution (i.e., $p_0(\X)$ no longer \textit{is} the prior, but
\textit{defines} the prior). We note that other measures of the
difference between distributions could be used to regularize the
Bayesian refinement.  The relative entropy provides us with a
positive-definite measure of deviation between $p(\X)$ and the
reference distribution $p_0(\X)$.  By weighting these deviations
exponentially, we arrive at a prior functional
\begin{equation}
  \label{eq:8a}
  {\cal P}_0[p(\X)] \propto \exp\left(-\theta \int d\X\, p(\X)\ln\frac{p(\X)}{p_0(\X)}\right),
\end{equation}
with a parameter $\theta>0$ expressing the level of confidence in
the reference ensemble, and therefore in the underlying potential
energy surface (force field) and the exhaustiveness of our sampling of
$p_0(\X)$. High confidence is expressed through large values of
$\theta$.  The choice of the confidence factor $\theta$ will be
discussed in the section on \textit{Practical Considerations} below.
Here and in the following, we use a calligraphic font for functionals,
and square brackets for their arguments.  The posterior functional
then becomes
\begin{eqnarray}
  \label{eq:9a}
  \lefteqn{{\cal P}[p(\X)|\mathrm{data}] }\nonumber\\&\propto&
  \exp\left(-\theta \int d\X\,
    p(\X)\ln\frac{p(\X)}{p_0(\X)}\right) {\cal P}[\mathrm{data}|p(\X)].
\end{eqnarray}
In the following, we will first consider the case where the measured
observables are properties common to all configurations before
considering the case where the observables are ensemble averages.

\paragraph{Ensemble refinement for properties common to all
  configurations.}

In some cases, ensemble refinement should be used even if the
observables are properties of individual configurations.  As an
example, consider a disulfide bond or other chemical cross-link that
is present in essentially all proteins within a system.  With respect
to other degrees of freedom, the configurations may be disordered.
Such cases require ensemble refinement, but with an experimental
restraint that acts on each ensemble member individually.

To quantify deviations from the observations, we use an approximate
likelihood functional.  For given $p(\X)$ and Gaussian errors
$\sigma_i$, the probability of the data is proportional to $\prod_i
\int d\X\,p(\X)\exp(-[y_i(\X)-y_i^{\OBS}]^2/2\sigma_i^2) \approx
\exp(-\int d\X\,p(\X)\sum_i[y_i(\X)-y_i^{\OBS}]^2/2\sigma_i^2)$,
ignoring higher-order fluctuations in the squared errors.  With this
approximation, we arrive at
\begin{equation}
  \label{eq:10a}
  {\cal P}[\mathrm{data}|p(\X)] = e^{-\chi^2[p(\X)]/2},
\end{equation}
where
\begin{equation}
  \label{eq:11a}
  \chi^2[p(\X)] = \sum_i \int d\X\,p(\X)
  \frac{\left[y_i(\X) -  y_i^{\OBS}\right]^2} {\sigma_i^2}
\end{equation}
is the mean-squared error of the common observables
$y_i(\X)$, scaled by $1/\sigma_i^2$.

To make progress, we now determine the normalized probability density
$\Popt(\X)$ that maximizes the posterior functional ${\cal P}[p(\X)|\Data]$. We define
\begin{eqnarray}
  \label{eq:5}
  \lefteqn{
    {\cal L}[p(\X)] \equiv -\ln {\cal P}[p(\X)|\Data] + \lambda \int d\X\,
    p(\X)} \nonumber\\
  &=&{\theta\int d\X\, p(\X)\ln\frac{p(\X)}{p_0(\X)}}\\&&  + \sum_i \int
  d\X\, p(\X) \frac{\left[y_i(\X)-y_i^{\OBS}\right]^2}
  {2\sigma_i^2}
  + \lambda \int d\X\, p(\X), \nonumber
\end{eqnarray}
where the Lagrange multiplier $\lambda$ is used to ensure
normalization, $\int d\X\,p(\X)=1$.  ${\cal L}$ trades off deviations
of $p(\X)$ from the reference distribution against deviations between
the predicted and measured observables.  Setting the functional
derivative with respect to $p(\X)$ to zero results in
\begin{eqnarray}
  \label{eq:6}
  \lefteqn{\frac{\delta{\cal L}}{\delta p(\X)} =\theta \left[
    \ln\frac{p(\X)}{p_0(\X)} + 1\right]}\nonumber \\ && + \sum_i
  \frac{\left[y_i(\X)-y_i^{\OBS}\right]^2}
  {2\sigma_i^2}+\lambda =0.
\end{eqnarray}
By solving this equation for $p(\X)\equiv\Popt(\X)$, we obtain an explicit expression
for the optimal probability density in common-property ensemble
refinement,
\begin{equation}
  \label{eq:7}
  \Popt(\X)\propto p_0(\X)\exp\left(-\sum_i
   \frac{\left[y_i(\X)-y_i^{\OBS}\right]^2}
   {2\theta\sigma_i^2}\right),
\end{equation}
which can then be normalized to one by integration over $\X$. 
We note that for $\theta=1$, this distribution is identical to
the Bayesian posterior of single-copy refinement in
Eqs.~(\ref{eq:1}-\ref{eq:3}).  

This procedure is closely related to the maximum-entropy method.  In
typical maximum-entropy approaches, measurements are imposed as strict
constraints.  By contrast, in the maximum-entropy formalism of Gull
and Daniell,\cite{Gull:78} noise is taken into account through a
$\chi^2$ term. However, $\chi^2$ enters in the form of a constraint to
match exactly an ``expected value'', and $\theta$ is the corresponding
Lagrange multiplier enforcing this constraint. Here, by contrast, we
have no \textit{a priori} expectations concerning the exact $\chi^2$
to be achieved in refinement.  Instead, we express our confidence in
the reference distribution through the choice of $\theta$ (even though
in practice, $\theta$ may be adjusted; see below).  We note that later
maximum entropy approaches accounting for noise in the data do not
always draw this distinction
\cite{Jaynes:1982,MaxEntNumericalRecipes,Rozycki:11} and minimize
functionals similar or identical to ${\cal L}$.

\paragraph{Refinement using ensemble averages.}

Next we assume that the measured quantities $Y_i^{\OBS}$ are averages
of observables $\YiEns_i(\X)$ over an ensemble of structures, as
represented by the functional
\begin{equation}
  \label{eq:43}
   {\cal Y}_i[p(\X)] = \int d\X\,p(\X)\YiEns_i(\X).
\end{equation}
For simplicity and concreteness, we assume Gaussian errors and a
likelihood functional correspondingly defined as
\begin{equation}
  \label{eq:10}
  {\cal P}[\mathrm{data}|p(\X)] = e^{-\chi^2[p(\X)]/2},
\end{equation}
where
\begin{equation}
  \label{eq:11}
  \chi^2[p(\X)] = \sum_i \frac{\left[\int d\X\,p(\X)\YiEns_i(\X) -
    Y_i^{\OBS}\right]^2} {\sigma_i^2}
\end{equation}
for a set of measurements $i$ of ensemble-averaged observables
$\YiEns_i(\X)$. For correlated errors, Eq.~(\ref{eq:4}) becomes
\begin{equation}
  \label{eq:12}
  \chi^2[p(\X)] = \mathbf{\delta Y}^T \bm{\Sigma}^{-1} \mathbf{\delta Y}
\end{equation}
with $\delta Y_i = \int d\X\,p(\X)\YiEns_i(\X)-Y_i^{\OBS}$.  We note
that the general formalism is of course not limited to Gaussian
errors.  Substituting $-2\ln {\cal P}[\mathrm{data}|p(\X)]$ for
$\chi^2$ will lead to the corresponding expressions for more general
likelihood functions.  We note further that more general functionals
can arise, e.g., if the measurements ${\cal Y}_i[p(\X)]$ report on
functions of averages, with measurements of the variance as the
simplest case.

In practice, one also has to deal with uncertainties
$\sigma_{i,\mathrm{calc}}^2$ in the forward calculation of the
observables $\YiEns_i(\X)$ from individual configurations $\X$.  Such
uncertainties often exceed the statistical errors
$\sigma_{i,\mathrm{obs}}^2$ in the measurements.  Assuming that the
two are uncorrelated, they can be lumped together, $\sigma_i^2 =
\sigma_{i,\mathrm{calc}}^2 + \sigma_{i,\mathrm{obs}}^2$.  Finally,
both errors can only be estimated with some uncertainty.  In a
Bayesian formulation, errors can be treated as nuisance parameters and
integrated out.\cite{Rieping:05}

\paragraph{Sampling from the Bayesian posterior functional in
  ensemble refinement.}
The above formulation appears to be of limited practical value, as one
would have to sample in function space.  One possible way to perform
such sampling in practice is to discretize the problem.  For instance,
clustering can be used to break up the configuration space into
discrete subsets.  If a set of configurations is drawn from the
reference distribution $p_0(\X)$, the relative weight $w_\alpha^0$ of
each cluster $\alpha$ would then be proportional to the number of its
members.  For cluster $\alpha$, the value for the observable $i$ is
$\YiEns_i^\alpha$, such that Eq.~(\ref{eq:11}) becomes
\begin{equation}
  \label{eq:13}
  \chi^2[w_1,w_2,\ldots,w_N] = \sum_i
  \frac{\left(\sum_{\alpha=1}^N w_\alpha \YiEns_i^\alpha-Y_i^{\OBS}\right)^2} {\sigma_i^2}
\end{equation}
with normalized weights $w_\alpha$.  These weights could then be
sampled according to
\begin{eqnarray}
  \label{eq:14}
  \lefteqn{
    {\cal P}[w_1,w_2,\ldots,w_N|\DATA] \propto }\\
  &&\exp\left(-\theta \sum_\alpha w_\alpha
    \ln \frac{w_\alpha}{w_\alpha^0}-\sum_i
    \frac{\left(\sum_{\alpha}w_\alpha \YiEns_i^\alpha-Y_i^{\OBS}\right)^2}
    {2\sigma_i^2}\right), \nonumber
\end{eqnarray}
again under the normalization constraint, $\sum_\alpha
w_\alpha=1$. Equation~(\ref{eq:14}) is the discrete analog of Eq.~(\ref{eq:9a}).
This form of Bayesian ensemble refinement can also be applied to a
collection of $N$ individual configurations, without clustering. If
one starts from an equilibrium ensemble of $\X_\alpha$ drawn from the
reference distribution $p_0(\X)$, then $w_\alpha^0=1/N$.

\paragraph{Optimal configuration space distribution from Bayesian
  ensemble reweighting.}
Instead of sampling the probability densities $p(\X)$ or
$\{w_1,w_2,\ldots,w_N\}$ from the posterior functional, we can again
try to find the most probable $p(\X)$ or $w_\alpha$, as in
Eqs.~(\ref{eq:5}-\ref{eq:7}) above. Configurations $\X$ sampled
according to this optimal $\Popt(\X)$ define representative ensembles.
To find the extremum of the posterior functional, we follow the same
variational approach as above and maximize the posterior functional in
Eq.~(\ref{eq:9a}) with respect to the probability density $p(\X)$.  As
optimization function, we use the negative logarithm of the posterior
${\cal P}$, with a Lagrange multiplier $\lambda$ to enforce
normalization.  For Gaussian errors, we obtain
\begin{eqnarray}
  \label{eq:15}
  \lefteqn{{\cal L}[p(\X)]=\theta \int d\X\, p(\X)\ln\frac{p(\X)}{p_0(\X)}}\\&& +
  \sum_i \frac{\left[\int d\X\, p(\X)\YiEns_i(\X) -Y_i^{\OBS}\right]^2}
  {2\sigma_i^2} +\lambda\int d\X\, p(\X),\nonumber
\end{eqnarray}
taking on a form that has been postulated as a starting point for a
maximum entropy approach.\cite{Gull:78,Rozycki:11} Here,
Eq.~(\ref{eq:15}) is a direct consequence of posterior maximization,
which would allow us to obtain corresponding log-posteriors also for more
non-Kullback-Leibler priors and more complicated likelihood functions
[e.g., for rigorous common-property refinement without the
approximation preceding Eq.~(\ref{eq:10a})]. In such cases,
postulating a proper maximum entropy formulation can be
difficult. Variational optimization of ${\cal L}$ results in
\begin{eqnarray}
  \label{eq:16}
  \lefteqn{\frac{\delta{\cal L}}{\delta p(\X)} =\theta \left[
    \ln\frac{p(\X)}{p_0(\X)} + 1\right]}\\ && + \sum_i
  \frac{\YiEns_i(\X)[\int d\X' p(\X')\YiEns_i(\X')-Y_i^{\OBS}]}
  {\sigma_i^2}+\lambda =0, \nonumber 
\end{eqnarray}
which can be solved formally to give
\begin{eqnarray}
  \label{eq:34}
  \lefteqn{\Popt(\X) \propto}\\
  && p_0(\X)\exp{\left[-\sum_i
      \frac{\YiEns_i(\X)\left[\int d\X' \Popt(\X')\YiEns_i(\X')-Y_i^{\OBS}\right]}
      {\theta\sigma_i^2}\right]}. \nonumber
\end{eqnarray}
We recognize Eq. (8) of Ref. \onlinecite{Boomsma:14:1}, albeit with a
somewhat different interpretation.  There, $1/\theta$ appears as a
Lagrange multiplier ``$\lambda$'' that has to be determined
self-consistently such that the $\chi^2$ for $\Popt(\X)$ matches a
desired value, following the maximum-entropy prescription of Gull and
Daniell;\cite{Gull:78} here, $\theta$ is a parameter that expresses a
priori the confidence in the reference distribution.  The
normalization factor in Eq. (\ref{eq:34}) can be determined by
integration [which is equivalent to determining our Lagrange
multiplier $\lambda$ in Eq.~(\ref{eq:16})].  For correlated errors of
the ensemble averages, Eq.~(\ref{eq:12}), the exponent in
Eq.~(\ref{eq:34}) should be replaced by
\begin{equation}
  \label{eq:34b}
  -\frac{1}{\theta}\sum_{i,j}\YiEns_i(\X) (\bm{\Sigma}^{-1})_{ij}\left[\int d\X'
    \Popt(\X')\YiEns_j(\X')-Y_j^{\OBS}\right].
\end{equation}
Because the weight function $\Popt(\X)$ appears inside the square in
the $\chi^2$ term of Eq.~(\ref{eq:15}), we have ended up with a
nonlinear integral equation, Eq.~(\ref{eq:34}), for $\Popt(\X)$ that
will usually be difficult to solve, in particular for high-dimensional
problems.  We note, however, that for refinement without explicit
consideration of errors, adaptive methods have been
developed.\cite{Pitera:12,White:14,White:15} Uncertainties are
considered by Beauchamp et al.,\cite{Beauchamp:14} albeit with a
number of additional priors introduced for constants acting as weight
factors in their bias.  In Appendix \ref{app:A}, we introduce an
adaptive algorithm to sample configurations according to the optimal
Bayesian ensemble distribution, Eqs.~(\ref{eq:34}) and (\ref{eq:34b}),
without the need of multiple replicas.

The above procedure can also be applied to problems with a set of $N$
discrete configurations.  We determine their optimal weights $\Wopt$,
$\alpha=1,\ldots,N$, by maximizing the negative log-posterior
\begin{eqnarray}
  \label{eq:17}
  \lefteqn{{\cal L}(w_1,\ldots,w_N)=\theta \sum_\alpha w_\alpha
    \ln \frac{w_\alpha}{w_\alpha^0}}\\
    && +\sum_i
    \frac{\left(\sum_{\alpha}w_\alpha \YiEns_i(\X_\alpha)-Y_i^{\OBS}\right)^2}
    {2\sigma_i^2}+ \lambda \sum_\alpha w_\alpha. \nonumber
\end{eqnarray}
The extremum of this negative log-posterior satisfies the following
set of coupled nonlinear equations
\begin{eqnarray}
  \label{eq:18}
  \lefteqn{\Wopt \propto }\\
  && w_\alpha^0\exp\left[-\sum_i
    \frac{\YiEns_i(\X_\alpha)\left(\sum_{\gamma}\WoptG \YiEns_i(\X_\gamma)-Y_i^{\OBS}\right)}
    {\theta\sigma_i^2}\right],\nonumber
\end{eqnarray}
which can be solved, for instance, by iteration, starting from
$w_\alpha^0$ (see below).  Alternatively, one can use simulated annealing or other
optimization methods to locate the global minimum of ${\cal L}$.

We note that the resulting optimal weights $\Wopt$ coincide exactly
with the EROS weights \cite{Rozycki:11} if all $N$ configurations are
reweighted.  In an illustrative example below, we will also consider
the case where sets of $n$ configurations are drawn from $p_0(\X)$ and
reweighted according to EROS.  In the limit of $n\rightarrow\infty$,
each structure enters this starting ensemble with the correct relative
weight. After EROS reweighting, using Eq.~(\ref{eq:18}) with
$w_\alpha^0=1/n$, one thus converges to the optimal Bayesian ensemble
refinement weights $\Wopt$ for $n\rightarrow\infty$.  This convergence
will be illustrated in a numerical example.

\subsection{Bayesian ensemble refinement in configuration space: The replica method}

The optimal weights $\Wopt$ determined self-consistently from
Eq. (\ref{eq:18}) can be used for reweighting of an ensemble of
structures drawn from the reference distribution. However, we cannot
use these weights directly to sample the ensemble of configurations on
the fly, lacking explicit solutions of Eqs.~(\ref{eq:34}) and (\ref{eq:18}) (but
see the Appendix for an adaptive method).

To circumvent the problem, we adopt a replica-based approach in which
averaged observables are calculated over multiple copies of the
system.\cite{Kim:89,Scheek:91,Kuriyan:91,Best:JACS:2004,Lindorff-Larsen:05}
In the replica simulations, $N$ copies (replicas) $\X_\alpha$ of a
molecular system are simulated in parallel using the same energy
function $U(\X_\alpha)$, subject in addition to a biasing potential
that attempts to match the observables obtained by averaging over the
$N$ copies to the experimental measurements.  We require that for a
single replica, $N=1$, one recovers the result of common-property
ensemble refinement, Eq.~(\ref{eq:7}).  At the other extreme,
$N\rightarrow\infty$, we want to recover the optimal Bayesian
configuration space distribution, Eq.~(\ref{eq:34}).

We use $N$ equally weighted replicas $\X_1,\ldots,\X_N$ to define a
function space of realizable probability densities,
$p(\X)=N^{-1}\sum_{\alpha=1}^N\delta(\X-\X_\alpha)$, where $\delta(x)$
is Dirac's delta function. To determine the relative weight of these
$p(\X)$, and in turn of the underlying replica states $\{\X_\alpha\}$,
we use the posterior functional Eq.~(\ref{eq:9a}) with the likelihood
in Eq.~(\ref{eq:10}),
\begin{eqnarray}
  \label{eq:19}
  \lefteqn{{\cal P}[p(\X)|\mathrm{data}]\propto e^{-\theta\int
    d\X\,p(\X)\ln\frac{p(\X)}{p_0(\X)}-\chi^2/2}}\nonumber\\
  &&\propto e^{\frac{\theta}{N}\sum_\alpha \ln
    p_0(\X_\alpha)-\chi^2/2 }\\
  &&=\prod_\alpha \left[p_0(\X_\alpha)\right]^{\frac{\theta}{N}}
  e^{-\sum_{i=1}^M\left[\frac{1}{N}\sum_{\gamma=1}^N 
      y_i(\X_\gamma)-Y_i^{\OBS}\right]^2/2\sigma_i^2}.\nonumber
\end{eqnarray}
For the evaluation of the entropy integral in the exponent, we
coarse-grained $\delta(x)$ as $1/\Delta$ for $|x|<\Delta/2$ and 0
otherwise; divided out a term proportional to $\ln\Delta$ because we
only require relative posterior probabilities; and then took the limit
$\Delta\to 0$.  Having chosen $N$-replica distributions as function
space, the $p(\X)$ are now parametrized by $\{\X_\alpha\}$, and in
Eq.~(\ref{eq:19}) the posterior functional has become a function that
can be interpreted as the sampling distribution of the replica states
$\{\X_\alpha\}$.  Here, we are interested in sampling from the
extremum of the posterior functional, i.e., the optimal Bayesian
ensemble distribution. To suppress fluctuations around the extremum as
$N\to\infty$, we take the posterior function in Eq.~(\ref{eq:19}) to a
power growing with $N$.  Taking it to the power $N/\theta$, we arrive
at the replica sampling distribution
\begin{eqnarray}
  \label{eq:21}
  \lefteqn{p_N(\X_1,\X_2,\ldots,\X_N) \propto \prod_{\alpha=1}^N p_0(\X_\alpha)\exp(-N\chi^2/2\theta)=} \\
  &&\exp\!\left[-\beta\sum_{\alpha=1}^N
    U(\X_\alpha) - \frac{N}{2} \sum_i
    \frac{\left[\frac{\sum_{\alpha=1}^N \YiEns_i(\X_\alpha)}{N}-Y_i^{\OBS}\right]^2}
    {\theta\sigma_i^2}\right]\!,\nonumber
\end{eqnarray}
with the Boltzmann factor for potential energy $U(\X)$ defining the
reference distribution. The second term in the exponent defines the
biasing potential applied to the ensemble of replicas.

We now show that under Eq.~(\ref{eq:21}) in the limit
$N\rightarrow\infty$, individual replicas indeed sample configurations
according to the optimal Bayesian ensemble refinement distribution in
Eq.~(\ref{eq:34}).  Without loss of generality, we determine the
distribution of replica 1, since all replicas are equivalent.  To this
end, we rewrite the last term in the exponent of Eq.~(\ref{eq:21}) as
\begin{eqnarray}
  \label{eq:35}
  \lefteqn{\frac{N}{2} \sum_i
  \frac{\left[\frac{1}{N}\sum_{\alpha=1}^N \YiEns_i(\X_\alpha)-Y_i^{\OBS}\right]^2}
  {\theta\sigma_i^2}}\nonumber\\
&=& \sum_i\left[
  \frac{\YiEns_i^2(\X_1)}{2\theta\sigma_i^2N}+\frac{\YiEns_i(\X_1)}{\theta\sigma_i^2}
  \left(\frac{1}{N}\sum_{\alpha=2}^N
    \YiEns_i(\X_\alpha)-Y_i^{\OBS}\right)\right]\nonumber\\
&&+\sum_i\frac{N\left(\frac{1}{N}\sum_{\alpha=2}^N
    \YiEns_i(\X_\alpha)-Y_i^{\OBS}\right)^2}{2\theta\sigma_i^2}.
\end{eqnarray}
Since the first term on the right is of order ${\cal O}(1/N)$ and the
second term is of order ${\cal O}(1)$, the first term vanishes in the
limit of $N\rightarrow\infty$.  In this limit, we can use a mean field
approximation for the second term, $\sum_{\alpha=2}^N
\YiEns_i(\X_\alpha)/N\approx\int d\X\,p(\X) \YiEns_i(\X)$.  The last
term on the right of Eq.~(\ref{eq:35}) is independent of $x_1$ and
thus cancels in the normalization of the resulting distribution over $\X_1$.  In
the limit of $N\rightarrow\infty$, we thus arrive at a
probability density for replica 1 (and, by symmetry, for all others)
of
\begin{eqnarray}
  \label{eq:36}
  \lefteqn{
  p(\X_1)\propto}\\&& p_0(\X_1) \exp{\left[-\sum_i
      \frac{\YiEns_i(\X_1)\left[\int d\X' p(\X')\YiEns_i(\X')-Y_i^{\OBS}\right]}
      {\theta\sigma_i^2}\right]},\nonumber
\end{eqnarray}
which is indeed identical to the probability density of optimal
Bayesian ensemble refinement in configuration space,
Eq.~(\ref{eq:34}).  Below, this identity will be demonstrated
explicitly for two analytically tractable models, and for a numerical
model.

Equation (\ref{eq:21}) for the probability density in Bayesian replica
ensemble refinement is nearly identical to that obtained by Cavalli et
al.\cite{Cavalli:13:1,Cavalli:Erratum:13} as a weighted integral over
the maximum entropy solution with strict constraints on the
observables.  However, there is one crucial difference: their $\chi^2$
in the exponent of the reweighting factor is missing the factor $N$
scaling the biasing potential with the number of replicas. Not scaling
$\chi^2$ by $N$ would result in decoupling of the replicas, as shown
explicitly below.  Indeed, early replica ensemble-refinement
simulations introduced the scale factor $N$
empirically,\cite{Best:JACS:2004} and it appears in a recent
preprint\cite{Bonomi:15} released shortly after submission of this
paper and release of a preprint.

Roux and Weare \cite{RouxWeare:13} also considered a maximum entropy
approach with strict constraints on the ensemble averages.  In
addition, these authors examined the convergence behavior of
$N$-replica simulations. For the specific example of a Gaussian
reference distribution and a harmonic restraint on the mean, Roux and
Weare \cite{RouxWeare:13} found that to recover the mean exactly for
large $N$, the effective spring constant in the biasing potential had
to grow faster than linearly in $N$.  For the general case, the choice
of the spring constant was left open. In our Bayesian formulation, we
account for the uncertainties of the measured averages.  It is
therefore not to be expected that the measurements are satisfied
strictly in the refined ensemble.  This will be illustrated below by
the analytical solution for the analogous problem of a Gaussian
reference distribution within our Bayesian framework.  More generally,
the explicit accounting for errors $\sigma_i$ provides a basis for
combining different measurements in a properly balanced manner.

On the basis of the preceding analysis, we note that if the $\chi^2$
term in Eq.~(\ref{eq:21}) were scaled by $N^a$ instead of $N$, with
$a>0$, then replica ensemble refinement would exhibit a ``phase
transition'' as a function of the exponent $a$ in the ``thermodynamic
limit'' of infinitely many replicas, $N\rightarrow\infty$.  For
sub-linear scaling, $0<a<1$, the effect of the $\chi^2$ bias vanishes
with increasing $N$ and the replica ensemble gradually falls back to the
reference distribution; for super-linear scaling, $a>1$, the $\chi^2$
bias diverges to infinity everywhere except at states that satisfy the
constraints exactly, making it equivalent to a sum of delta functions
that impose strict constraints on the averages; only for linear
scaling, $a=1$, replica sampling converges to the distribution of
optimal Bayesian ensemble refinement.  This $N$-scaling becomes
explicit in the Gaussian models studied by Roux and Weare
\cite{RouxWeare:13} and below.

\subsection{BioEn method combining replica simulations with EROS}

In the following, we describe the BioEn algorithm that simultaneously
addresses the possible shortcomings of EROS and Bayesian replica
ensemble refinement and helps us in the choice of the $\theta$
parameter.  By combining EROS and replica simulations, one can
accelerate the convergence toward the optimal Bayesian ensemble.  This
combination also makes it possible to obtain optimal Bayesian
ensembles for a wide range of $\theta$ values without the need to run
actual Bayesian replica simulations for all of them.  Covering a broad
$\theta$ range is important in practice to choose a suitable
confidence parameter $\theta$ that achieves a good balance between
reference distribution and data.

In EROS, one can work with large numbers $n$ of structures without
significant computational costs; however, if $p_0(\X)$ and $\Popt(\X)$
have little overlap in configuration space, then these structures may
not be representative of the refined ensemble, resulting in slow
convergence with increasing $n$, as shown below.  By contrast, the
computational cost of sampling Bayesian replica ensembles with large
$N$ is high.  To accelerate the convergence toward $N$-independent
optimal Bayesian ensemble refinement, one can combine the replica and
EROS methods (and the adaptive method described in the Appendix). 
Even with relatively small $N$, the replica simulations
can be used to enrich the sample of configurations fed
into EROS refinement.  To give these configurations $\X$ the proper
weight proportional to $p_0(\X)$, with $\X$ coming from different
simulations with and without bias, one can for instance use a
histogram-free version of the multidimensional weighted histogram
analysis method (WHAM).\cite{Rosta:JACS:2011,Shirts:08,Souaille:01}

We first need to reweight the $N$-replica states in different
simulations according to the reference distribution.  By running $N$ unbiased,
uncoupled simulations according to $p_0(\X_1)p_0(\X_2)\cdots
p_0(\X_N)$ (or, simply, one unbiased simulation as a source for
configurations $\X_\alpha$ that are then combined at random to form
pseudo $N$-replica states) and one or several biased, coupled
simulations (e.g., with different $\theta_i$) according to
$p_N(\X_1,\ldots,\X_N)$, one obtains representative sets of
$N$-replicas states. To combine them, one needs to assign the proper
relative weight $w_{i,k}^0$ to the $k$-th sampled $N$-replica state
$\{\X_\alpha\}_{i,k}$ in run $i$, as given by the reference distribution
$\prod_{\alpha=1}^N p_0(\X_\alpha)$.  Following
Ref. \onlinecite{Rosta:JACS:2011}, we first determine the free
energies $F_i$ of each $N$-replica simulation $i$ ($i=1,\ldots,M_\mathrm{run}$) by
iteratively solving the coupled set of equations
\begin{equation}
  \label{eq:41}
  e^{-\beta F_i} = \sum_{m=1}^{M_\mathrm{run}} \sum_{k=1}^{n_m}\frac{e^{-\beta
      U_i(\{\X_\alpha\}_{m,k})}} {\sum_{j=1}^{M_\mathrm{run}} n_j e^{\beta[F_j - 
      U_j(\{\X_\alpha\}_{m,k})]}},
\end{equation}
where $F_1\equiv 0$ by definition.  The outer sums on the right extend
over the ${M_\mathrm{run}}$ runs (indexed by $m$ and $j$) and the $n_m$ $N$-replica
states (indexed by $k$) in run $m$.  The biasing potential is defined
as $U_i(\{\X_\alpha\})\equiv N\chi^2/2\theta_i$ in biased runs $i$,
and $U_i\equiv 0$ in unbiased ones, with $\chi^2$ as in
Eq.~(\ref{eq:21}).  To obtain the relative weight $w_{i,k}^0$ of
replica state $k$ in run $i$, $\{\X_\alpha\}_{i,k}$, corresponding to
the reference distribution, we set the $\delta$-term in Eq.~(2) of
Ref. \onlinecite{Rosta:JACS:2011} equal to one for only this replica
state and to zero for all others.  We then obtain
\begin{equation}
  \label{eq:42}
  w_{i,k}^0 \propto \left[\sum_{j=1}^{M_\mathrm{run}} n_j
  e^{\beta [F_j-U_j(\{\X_\alpha\}_{i,k})]}\right]^{-1},   
\end{equation}
where the sum extends over the different simulations $j$.  Each of the
$N$ configurations $\X_\alpha$ in a given replica state
$\{\X_\alpha\}_{i,k}$ then has the same relative weight $w_{i,k}^0$ as
the replica state.  The resulting set of configurations together with
their estimated relative weights can then be used as input for an EROS
refinement according to Eq.~(\ref{eq:18}).  With the resulting
EROS-refined weights, one obtains the BioEn ensemble of configurations
enriched by the biased $N$-replica simulations, yet properly
reweighted to correct for effects of finite numbers $N$ of replicas.
Importantly, this reweighting approach also allows one to obtain
optimally reweighted ensembles for different $\theta$, simply by
re-running EROS.

\section{Results and Discussion}

\subsection{Illustrative examples of ensemble reweighting}

\paragraph{Ensemble reweighting of the mean.}

To illustrate and test the ensemble reweighting formalisms described
above, we consider a simple, analytically tractable problem.  Consider
a one-dimensional configuration coordinate $x$ with a Gaussian
reference distribution
\begin{equation}
  \label{eq:22}
  p_0(x)={\left(2\pi s^2\right)^{-1/2}}{e^{-{x^2}/{2s^2}}}
\end{equation}
and the mean $Y=\overline{x}$ as observable, with uncertainty
$\sigma$, such that
\begin{equation}
  \label{eq:23}
  \chi^2=\frac{\left[\int dx\,p(x)x-Y\right]^2}{\sigma^2}.
\end{equation}
This problem is closely related to a Gaussian model for replica
simulations studied by Roux and Weare,\cite{RouxWeare:13} with the
difference that here we explicitly account for the uncertainty
$\sigma$ in the measured mean $Y$.  The negative log-posterior
Eq.~(\ref{eq:15}) becomes
\begin{eqnarray}
  \label{eq:24}
  {\cal L}[p(x)] & = & \theta\int dx\, p(x)\ln \frac{p(x)}{p_0(x)} +
  \frac{\left[\int dx\,p(x)x-Y\right]^2}{2\sigma^2}\nonumber\\&&
  + \lambda \int dx\,p(x).
\end{eqnarray}
The extremum of ${\cal L}$ satisfies
\begin{equation}
  \label{eq:25}
  \frac{\delta{\cal L}}{\delta p(x)} =\theta \left[
    \ln\frac{p(x)}{p_0(x)} + 1\right]+
  \frac{x\left[\int dx'\,p(x')x'-Y\right]}{\sigma^2}
    + \lambda
\end{equation}
This integral equation can be solved with a Gaussian ansatz,
$p(x)=(2\pi s^2)^{-1/2} \exp[-(x-\mu)^2/(2s^2)]$, with $\theta$ set to
one without loss of generality, since a change in $\theta$ here
corresponds to a rescaled $\sigma^2$ (see the Appendix for an
alternative solution method using generating functions).
By substituting the ansatz into
the integral equation and solving for the coefficients of powers of
$x$, it follows that the mean of the optimal probability density is
\begin{equation}
  \label{eq:8}
  \mu=Ys^2/(s^2+\sigma^2).
\end{equation}
The optimal probability density of Bayesian ensemble refinement is
thus a Gaussian,
\begin{equation}
  \label{eq:26}
  \Popt(x)={\left(2\pi s^2\right)^{-1/2}}
  {\exp\left[-\frac{\left(x-\frac{Ys^2}{s^2+\sigma^2}\right)^2}{2s^2}\right]}.
\end{equation}
The variance $s^2$ remains unchanged from the reference distribution,
but the mean is shifted from zero toward the ensemble average $Y$
according to the relative weights of the variances in the reference
distribution, $s^2$, and in the $\chi^2$ error, $\sigma^2$.  In the
limit $\sigma\ll s$, the mean approaches the measurement $Y$; in the
opposite limit $\sigma\gg s$, the mean remains near that of the
reference distribution, i.e., at zero.

Bayesian replica ensemble refinement for this problem is also
analytically tractable.  For $N$ replicas, with $\theta=1$, we have
\begin{equation}
  \label{eq:27}
  p_N(x_1,\ldots,x_N)\propto
\frac{\exp\left[-\frac{\sum_{\alpha=1}^Nx_\alpha^2}{2s^2}-\frac{N\left(N^{-1}\sum_\alpha x_\alpha -Y \right)^2}{2\sigma^2}\right]}{(2\pi s^2)^{N/2}}. 
\end{equation}
Since this replica probability density is symmetric in exchanges of the $x_i$,
all replicas sample the same space, and we can integrate out all $x_i$
but $x_1$ to obtain a marginalized replica probability density
\begin{equation}
  \label{eq:28}
  q_N(x_1)=\int dx_2\,dx_3\cdots dx_N p_N(x_1,\ldots,x_N).
\end{equation}
The Gaussian integrals can be carried out, resulting in $q_N(x_1)$
being Gaussian with mean $Ys^2/(s^2+\sigma^2)$ and variance
$s^2\left[1-s^2/N(s^2+\sigma^2)\right]$.
For $N=1$, we recover the probability of common-property refinement,
Eq.~(\ref{eq:7}), with variance $1/(s^{-2}+\sigma^{-2})$.  In the limit
of $N\rightarrow\infty$, the variance approaches $s^2$.  We thus have
$\lim_{N\rightarrow\infty}q_N(x)=\Popt(x)$, with $\Popt(x)$ the optimal
Bayesian ensemble refinement result in Eq.~(\ref{eq:26}). Series
expansion shows that this limit is approached asymptotically as
$\ln[q_N(x)/\Popt(x)]=f(x)/N+{\cal O}(1/N^2)$ with
$f(x)=(s^2-x^2)/2(s^2+\sigma^2)$ for fixed $x$, i.e., as ${\cal
  O}(1/N)$ in the error of the logarithm of the probability density.

Importantly, if we had left out the factor $N$ scaling the $\chi^2$ in
the exponent of Eq.~(\ref{eq:27}), the mean would instead have been
$Ys^2/(s^2+N\sigma^2)$.  The mean would thus approach zero, i.e., the
value of the reference distribution, as the number of replicas is
increased, $N\rightarrow\infty$, irrespective of the uncertainty
$\sigma>0$.  This result makes it clear that the $\chi^2$ in the
replica model has to be scaled by $N$ to obtain a result that is
size-consistent in the number of replicas $N$.

\paragraph{Ensemble reweighting of the second moment.}

Another analytically tractable ensemble reweighting problem is
obtained for a non-linear observable, the second moment
$Y=\overline{x^2}$, again for the Gaussian reference distribution in
Eq.~(\ref{eq:22}).  For the second moment, we have $\chi^2=\left[\int
  dx\, p(x)x^2-Y\right]^2/\sigma^2$.  The optimal solution then has to
satisfy the integral equation
\begin{equation}
  \label{eq:29}
  \frac{\delta{\cal L}}{\delta p(x)} =\theta \left[
    \ln\frac{p(x)}{p_0(x)} + 1\right]+
  \frac{x^2\left[\int dx'\,p(x')x'^2-Y\right]}{\sigma^2}
    + \lambda~.
\end{equation}
To solve this integral equation, we make a Gaussian ansatz with zero
mean and variance $t^2$, $p(x)=\exp[-x^2/(2t^2)]/(2\pi t^2)^{1/2}$,
with $\theta$ set to one without loss of generality, and find
\begin{equation}
  \label{eq:30}
  t^2=\frac{2Ys^2-\sigma^2+\left[8\sigma^2s^4+\left(\sigma^2-2Ys^2\right)^2\right]^{1/2}}
  {4s^2}.
\end{equation}
In the limit of no uncertainty in the measurement, $\sigma\rightarrow
0$, we find that $t^2\rightarrow Y$, i.e., we have a Gaussian with
exactly the measured second moment.  In the other limit of complete
uncertainty, $\sigma\rightarrow\infty$, we have $t^2=s^2$, i.e., no
change relative to the reference distribution.  In between, $t^2$ is a
nonlinear interpolation between these two extremes.

This problem is also analytically tractable for Bayesian replica
ensemble refinement.  For $N$ replicas with $\theta=\beta=1$, the
joint probability density is
\begin{eqnarray}
  \label{eq:31}
  \lefteqn{
  p_N(x_1,\ldots,x_N)\propto}\\&&
\exp\left[-\frac{\sum_{\alpha=1}^Nx_\alpha^2}{2s^2}
      -\frac{N}{2}\frac{\left(\frac{1}{N}\sum_\alpha x_\alpha^2 -Y \right)^2}{\sigma^2}\right].\nonumber
\end{eqnarray}
To integrate out $x_2$ to $x_N$, we introduce $(N-1)$-dimensional
spherical coordinates, with $r^2=\sum_{\alpha=2}^N x_\alpha^2$.  The
marginalized distribution then becomes
\begin{eqnarray}
  \label{eq:32}
  \lefteqn{q_N(x_1)=\int dx_2\,dx_3\cdots dx_N p_N(x_1,\ldots,x_N)}\\
    &\propto&\int dr\,r^{N-2}\exp\left[-\frac{r^2+x_1^2}{2s^2}
      -\frac{N}{2}\frac{\left(\frac{r^2+x_1^2}{N} -Y\right)^2}{\sigma^2}\right] \nonumber\\
    &\equiv & \int dr\,f(r|x_1)\nonumber
\end{eqnarray}
As it turns out, the remaining one-dimensional integral can be carried
out analytically, giving an expression in terms of confluent
hypergeometric functions.  However, to take the $N\rightarrow\infty$
limit, it is advantageous to use a saddle-point approximation of the
integrand in terms of a Gaussian, $f(r|x)\approx
f_0\exp[-(r-\mu)^2/2v]$, that becomes increasingly accurate as $N$
increases.  We find that the variance $v$ in $r$ becomes independent
of $N$ and $x$ in the limit of large $N$, such that only the value
$f_0=f(\mu|x)$ at the extremum needs to be considered in the
construction of the marginalized distribution of $x$.  In the limit of
$N\rightarrow\infty$, $f_0$ depends on $x$ as
\begin{equation}
  \label{eq:33}
  \ln f_0(x) \approx \ln f_0(x=0)-\frac{x^2}{2t^2},
\end{equation}
where the $x=0$ value cancels in the normalization of the marginalized
distribution $q_N(x)$.  The marginalized distribution of $x$ is thus a
Gaussian centered at zero with variance $t^2$, as given in
Eq.~(\ref{eq:30}).  The result of Bayesian replica ensemble refinement
thus converges to the probability density of optimal Bayesian ensemble
refinement in the limit of large $N$.  This correspondence once again
stresses the importance of scaling the $\chi^2$ by $N$ to maintain
proper coupling and convergence in the limit of large numbers $N$ of
replicas.

\begin{figure}[htb]
  \centering
  \includegraphics[width=8cm]{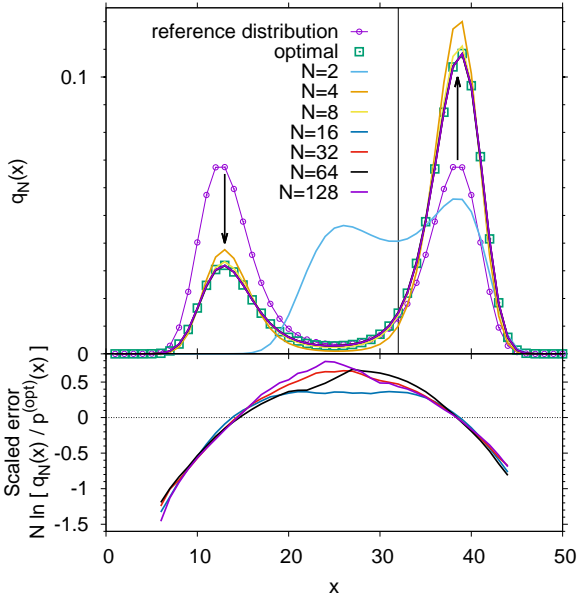}
  \caption{Optimal Bayesian ensemble refinement and Bayesian replica
    ensemble refinement for one-dimensional double well system with
    restraint on the mean, as indicated by the vertical black
    line. (Top) Marginalized distributions $q_N(x)$ from Bayesian
    replica ensemble refinement with $N$ replicas (lines) compared to
    the optimal Bayesian ensemble refinement solution $\Popt(x)$ (open
    squares) and to the reference distribution $p_0(x)$ (thin line with open
    circles). Arrows indicate changes relative to $p_0(x)$.  (Bottom)
    Error $N [\ln q_N(x)-\ln \Popt(x)]$ in $\ln q_N(x)$ scaled by the
    number of replicas $N$. Part of the scatter is a reflection of the
    stochastic Monte Carlo sampling of the Bayesian distributions
    $q_N(x)$.
    \label{fig:2}}
\end{figure}

\paragraph{Convergence of Bayesian replica ensemble refinement.}
To examine the convergence of the Bayesian replica ensemble refinement
with the number $N$ of replicas towards the optimal Bayesian ensemble
refinement solution, we have performed Metropolis Monte Carlo
simulations for a one-dimensional system defined by a double-well
potential energy function $\beta U(x)=3[(x-a)^2-b^2]^2/b^4$ with
$a=(M+1)/2$ and $b=(M+2)/4$. We have discretized the potential at
$x=1, 2, \ldots, M$ with $M=50$.  The mean of the corresponding
reference distribution $p_0(x)\propto\exp[-\beta U(x)]$ is at
$\overline{x}=a=25.5$. In the ensemble refinement, we set the target
half-way between the maximum and the upper minimum, at
$Y=(6+5M)/8=32$.  The resulting $\chi^2$ then becomes
$\chi^2=(N^{-1}\sum_{\alpha=1}^Mx_\alpha-Y)^2/\sigma^2$, with $\sigma$
set to one.  Systems with $N=2, 4, 8, \ldots, 128$ replicas were
sampled with Monte Carlo simulations, and the distributions $q_N(x)$
averaged over all replicas calculated.

Figure \ref{fig:2} compares the resulting distributions $q_N(x)$ of
$x$ from Bayesian replica ensemble refinement to $\Popt(x)$ from
optimal Bayesian ensemble refinement. We find that the ensemble
reweighted distributions shift contributions from the left well to the
right well to match the target mean, but by and large retain the shape
within each well of the potential $U(x)$ defining the reference
distribution.  The only exception is $N=2$, where the restraint on the
mean effectively pulls one of the replicas out of the first minimum
into the barrier region.  We also find numerically that the
distributions $q_N(x)$, averaged over all replicas $N$, converge
asymptotically (for large $N$) to the optimal Bayesian ensemble
refinement solution $\Popt(x)$ as $\ln q_N(x)\approx \ln \Popt(x) +
f(x)/N$ for $N\ge 16$.  Numerical results for the master curve
$f(x)=N\ln[q_N(x)/\Popt(x)]$ are shown in the bottom panel of Figure
\ref{fig:2} bottom; the actual error in an $N$-replica simulation is
approximately $1/N$-th of $f(x)$.  The probability density from
Bayesian replica ensemble refinement thus appears to converge
asymptotically as $1/N$ to the optimal Bayesian result, as in the
first analytically tractable example above.

\paragraph{Convergence of EROS.}

We have used the same model to examine the convergence of EROS in the
case where $n$ representative configurations are drawn according to
$p_0(x)$ and then reweighted according to Eq.~(\ref{eq:18}).
Specifically, we have drawn $n$ values of $x$ with replacement
according to the Boltzmann distribution for the double-well potential
with $M=50$.  The resulting $n$ points, indexed as
$\alpha(1),\ldots,\alpha(n)$, were then reweighted according to
Eq.~(\ref{eq:18}), with $w_{\alpha(i)}^0=1/n$ for all $i$.  The
resulting EROS weights were then averaged for each of the $M$ possible
values of $x$,
\begin{equation}
  \label{eq:40}
  p_n(x=x_\alpha)=\left\langle \sum_{i=1}^n
    \delta_{\alpha,\alpha(i)}w_{\alpha(i)}^{(\mathrm{opt})}\right\rangle,  
\end{equation}
where $\langle\cdots\rangle$ indicates an average over repeated
selections of samples of size $n$, and $\delta_{\alpha,\gamma}=1$ if
$\alpha=\gamma$ and zero otherwise.  In this way, we estimated the
expected weight of configuration $\alpha$ in repeated EROS runs using
$n$ representative ensembles.

\begin{figure}[htb]
  \centering
  \includegraphics[width=8cm]{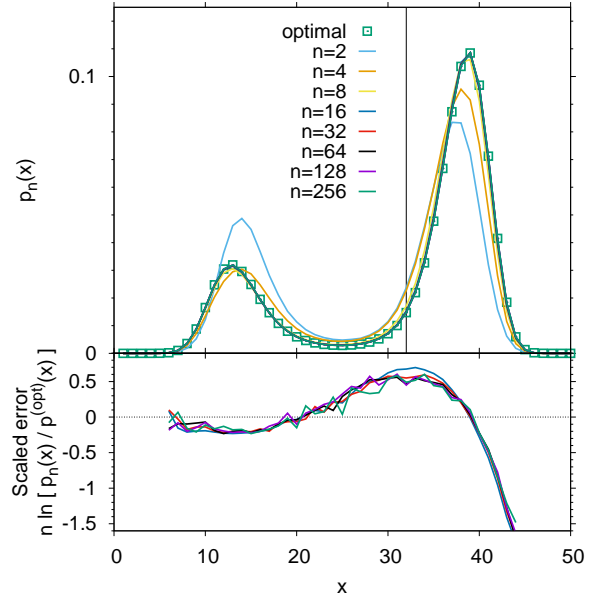}
  \caption{EROS and optimal Bayesian ensemble refinement for a
    one-dimensional double well system with restraint on the mean, as
    indicated by the vertical black line. (Top) Marginalized
    distributions $p_n(x)$ from EROS with $n$ configurations drawn
    according to $p_0$ (lines) compared to the optimal Bayesian ensemble
    refinement solution $\Popt(x)$ (open squares).  (Bottom) Error $n[\ln
    p_n(x)-\ln \Popt(x)]$ in $\ln p_n(x)$ scaled by the sample size
    $n$. Part of the scatter is a reflection of the stochastic Monte
    Carlo sampling of configurations in EROS.  \label{fig:3}}
\end{figure}

In Figure~\ref{fig:3}, we show that the distribution $p_n(x=x_\alpha)$
obtained by repeated reweighting indeed converges to the optimal
Bayesian ensemble refinement result in the limit of large $n$.  For
the specific example, the error in $\ln p_n(x)$ scales as
$1/n$. Interestingly, the relative error obtained for EROS samples of
size $n$ is comparable to that of Bayesian replica ensemble refinement
with $n$ replicas.

\paragraph{BioEn improves convergence by combining EROS and replica
  simulations.}

\begin{figure}[htb]
  \centering
  \includegraphics[width=8cm]{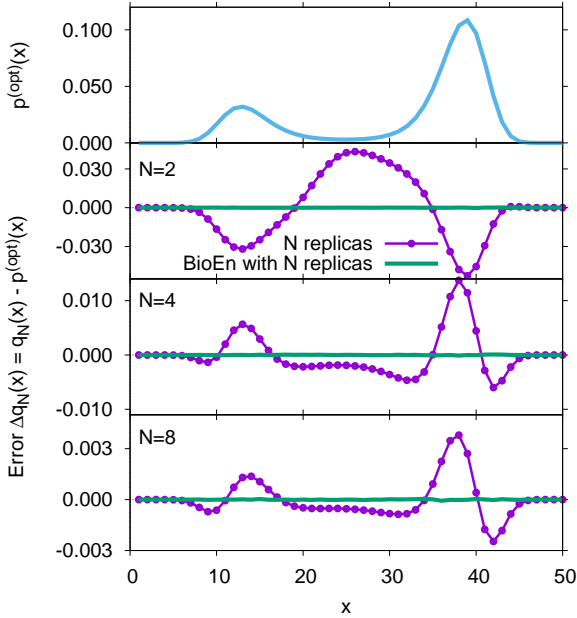}
  \caption{BioEn method applied to double-well system.  The bottom
    three panels show the error $\Delta q_N(x) = q_N(x) -\Popt(x)$ in
    the ensemble distributions obtained from regular replica ensemble
    refinement with $N=2$, 4, and 8 replicas (purple line with
    symbols) relative to the optimal Bayesian ensemble distribution
    $\Popt(x)$ (shown in the top panel). Also shown is the error of
    the BioEn method combining EROS and replica refinement (green
    lines).  Please note the change in scale of the vertical axes for
    different $N$.  See Figure~\protect\ref{fig:2} for the error in
    $\ln q_N(x)$ without BioEn for larger numbers of replicas
    $N$. \label{fig:4}}
\end{figure}

We have also tested the BioEn combination of EROS and replica
simulations to speed up convergence to the optimal Bayesian ensemble
distribution.  Figure~\ref{fig:4} demonstrates the dramatic
improvement achieved in the combined method for the model of
Figures~\ref{fig:2} and \ref{fig:3}.  After EROS reweighting of the
configurations sampled in unbiased and biased runs with $N=2$, 4, and
8 replicas, we find that the significant systematic errors in the
Bayesian replica ensemble distributions $q_N(x)$ disappear, and only
small, primarily statistical errors remain.

\subsection{Practical considerations}

\paragraph{Combining common-property and ensemble-average refinement.}

For simplicity, we have so far dealt separately with observables
reporting on properties common to all configurations and on ensemble
averages.  However, these data can be combined readily within the
above formalisms.  In the respective posteriors, the likelihood terms
according to Eqs.~(\ref{eq:10a}) and (\ref{eq:10}) simply have to be
multiplied.  The optimal Bayesian ensemble distribution then becomes
\begin{eqnarray}
  \label{eq:34a}
  \lefteqn{\Popt(\X) \propto  p_0(\X) \exp{\left[-\sum_{i=1}^{m}
        \frac{\left[y_i(\X)-y_i^{\OBS}\right]^2}
        {2\theta\sigma_i^2}\right]}}\\
  &&\times\exp{\left[-\sum_{i=m+1}^M
      \frac{\YiEns_i(\X)\left[\int d\X' \Popt(\X')\YiEns_i(\X')-Y_i^{\OBS}\right]}
      {\theta\sigma_i^2}\right]}, \nonumber
\end{eqnarray}
for $m$ restraints on common properties, and $M-m$ restraints on
ensemble averages.  Equation~(\ref{eq:34a}) is a combination of
Eqs.~(\ref{eq:7}) and (\ref{eq:34}).  An analogous expression
generalizes Eq.~(\ref{eq:18}) for the discrete case.

\paragraph{Data from single-molecule experiments.}

Data from single-molecule experiments can be incorporated in the
different refinement procedures.  In principle, one could even fit the
data individually, one molecule at a time, using single-copy
refinement.  In a more practical approach, one can use the techniques
of ensemble refinement to fit the single-molecule data lumped together
in a way that produces not just averages but also distributions of
observables.  An example are FRET efficiencies $E$ measured by
single-molecule spectroscopy. As a basis for ensemble
refinement,\cite{Boura:11,Boura:12} one can for instance determine
FRET-efficiency histograms $H_i^{\OBS}\equiv H^{\OBS}(E_i)$ from
photon arrival trajectories and use the deviations between histogram
counts calculated for an ensemble model, $H_i\equiv {\cal Y}_i[p(x)]$,
and measured in experiment, $H_i^{\OBS}\equiv Y_i^{\OBS}$, to
construct a $\chi^2$, with appropriate error models.  With such a
$\chi^2$, one can then use both EROS \cite{Boura:11,Boura:12} and
Bayesian replica ensemble refinement.

\paragraph{Dynamic, time-dependent data such as NMR NOEs.}
\label{sec:dynamics} 
Many relevant observables are not just functions of a configuration,
$y_i\equiv y_i(\X)$, but depend also on the dynamics.  Examples are
the NOE intensity and other NMR relaxation parameters that depend on
the rotational and translational dynamics of the spin
system.\cite{Peter:01} Whereas it is outside of the scope of this
article to refine an entire dynamical model to such data, we can make
some progress in this direction by considering a reduced problem.
Ignoring self-consistency issues, we can attempt to refine an ensemble
of configurations $\X$ that evolve in time under the Hamiltonian of
the molecular simulation energy function $U(\X)$ defining the
reference distribution $p_0(\X)$, but are distributed according to
$\Popt(\X)$ instead of $p_0(\X)$.

To calculate the observables associated with a particular
configuration $\X$, one can use trajectory segments passing through
$\X$.  Each of the sample configurations $\X$ would then serve as an
initial value, with Maxwell-Boltzmann velocities, for one or multiple
trajectory segments of length $\tau/2$.  To center the trajectories at
$\X$ with respect to time, one can run trajectory pairs of length
$\tau/2$, initiated from $\X$ with sign-inverted Maxwell-Boltzmann
velocities, one running forward and the other running backward in
time. Stitching the two segments together at $\X$, after
sign-inverting the velocities of the backward segment, one obtains a
continuous trajectory of length $\tau$ centered time-wise at $\X$.
For each of these trajectories, the time-dependent observable
$\YiEns_i=\YiEns_i[\X(t)|\X(0)=\X;-\tau/2\le t\le \tau/2]$ can be
calculated, possibly averaged by repeated runs over different choices
of initial Maxwell-Boltzmann velocities.  The $y_i$ calculated in this
manner can be treated as simple functions of $\X=\X(0)$ to enter the $\chi^2$
in the same way as static data.  The trajectory length $\tau$ should
be set such that the $y_i$ can be calculated with reasonable accuracy
(i.e., as multiples of the relevant correlation times).  After
refinement, one obtains an ensemble of configurations that jointly
account for the time-dependent observables yet stay close to the
reference distribution.  We note that (possibly overlapping)
trajectory segments could also be obtained from long equilibrium
trajectories, or even from $N$-replica simulations.

As the simplest approach of refining also the actual dynamics, one can
perform in addition time scaling, $t\mapsto\alpha t$, which could for
instance account for incorrect viscosities of the water model used in
the molecular dynamics simulations.  The time-scale parameter $\alpha$
can then be optimized as well in the ensemble refinement.

\paragraph{Solving the EROS equations.}

One can obtain the EROS weights in Eq.~(\ref{eq:18}) by numerical
minimization of ${\cal L}$ in Eq.~(\ref{eq:17}), which can be
accomplished by a variety of techniques with and without gradient
calculations.  Alternatively, one can solve Eq.~(\ref{eq:18})
directly, for instance by iteration until self-consistency is
achieved.  A possible route is to start from the weights
$\Wopt\approx w_\alpha^0$ in the reference distribution, and then
iterate Eq.~(\ref{eq:18}) to get an updated estimate of $\Wopt$.  This
procedure can be repeated until the change in old and new
approximations drops below a chosen threshold.  We found that mixing
the old and new approximations geometrically, as
${(w^{(\mathrm{old})}_\alpha)}^x{(w_\alpha^{(\mathrm{new})})}^{(1-x)}$
with $0<x<1$, led to stable fixed-point iterations.  The mixing
parameter $x$ controls stability ($x\approx 1$) and speed
($x\approx 0$).  We further improved the stability and convergence
behavior by starting at a large value of $\theta$, where the
deviations from the reference distribution $w_\alpha^0$ are small, and
then reducing $\theta$ in repeated fixed-point iterations to sweep out
a broad $\theta$ range.  The resulting EROS refinements for different
$\theta$ can help us in the choice of $\theta$, as discussed next.

\paragraph{Choosing the confidence factor $\theta$.}
\label{sec:theta}
The Bayesian ensemble refinement methods described here contain one
free parameter, the factor $\theta$ that enters the prior and
quantifies the level of confidence one has in the reference
probability density $p_0(\X)$.  Large values of $\theta$ express high
confidence (for instance, if one uses a well-tested atomistic force
field instead of a more approximate coarse-grained representation,
both being well sampled).  Whereas formally, one would choose $\theta$
before refinement, in practice one may want to readjust this choice
after the fact to achieve a better balance between reference
distribution and data.  By reporting the chosen $\theta$ and the
corresponding Kullback-Leibler divergence between reference and
optimal distribution, the inference process becomes transparent.

Akin to L-curve selection in other regularization approaches to
inverse problems,\cite{Hansen:93} one can find an appropriate value
of $\theta$ by plotting the Kullback-Leibler divergence (relative
entropy) $S_\mathrm{KL}=-\sum_\alpha \Wopt \ln({\Wopt}/{w_\alpha^0})$
against the $\chi^2$ obtained in EROS reweighting for different values
of $\theta$.  As discussed above, EROS reweighting can (and should) be
performed even when Bayesian replica ensemble refinement is used to
obtain the configurations to avoid finite-$N$ effects. The value of
$\beta^{-1}S_\mathrm{KL}$ can then be interpreted as the average error
in the energy function $U(\X)$ used to define the reference
distribution, since by definition $S_\mathrm{KL}=\beta\int
d\X\,\Popt(\X)\,[ U^{(\mathrm{opt})}(\X)-U(\X)]$ for $\Popt(\X)\propto
\exp[-\beta U^{(\mathrm{opt})}(\X)]$, given that the additive constant
in $U^{(\mathrm{opt})}(\X)$ is chosen such that the partition
functions (and thus free energies) of the optimal and reference
distribution are identical, $\int d\X \exp[-\beta
U^{(\mathrm{opt})}(\X)]=\int d\X \exp[-\beta U(\X)]$.  This allows one
to choose a $\theta$ value on the basis of expectations concerning the
magnitude of this error.\cite{Rozycki:11}  Conversely, one can also
take a more pragmatic approach and choose a value of $\theta$ at the
kink of the $S_\mathrm{KL}$-versus-$\chi^2$ curve, where a further
decrease in $\theta$ does not produce a significant improvement in the
fit quality but causes a large deviation from the reference
distribution, as measured by $S_\mathrm{KL}$.  This approach is taken
in the MERA web server for the refinement of peptide Ramachandran maps
against NMR data.\cite{Mantsyzov:14,Mantsyzov:15}  There one accepts
a $\chi^2$ a certain percentage point (say, 25 {\%}) above the minimal
$\chi^2$ obtained for $\theta\approx 0$.

Finally, the confidence parameter $\theta$ can also be treated as a
nuisance parameter with an uninformative prior, $p(\theta)\propto
1/\theta$ for $\theta>0$.  One could include $\theta$ in the
maximization of the posterior or attempt to integrate it out in a
weighted average over ensembles obtained for fixed $\theta$.

\section{Conclusions}

We have described different Bayesian approaches to ensemble
refinement, established their interrelations, and shown how they can
be applied to experimental data.  The Bayesian approaches allow one to
integrate a wide variety of experiments, including experiments
reporting on properties common to all configurations and on averages
over the entire ensemble.  We started from a Bayesian formulation in
which the posterior is a functional that ranks the quality of the
configurational distributions.  We then derived expressions for the
optimal probability distribution in configuration space. For discrete
configurations, we found that this optimal distribution is identical
to that obtained by the EROS model.\cite{Rozycki:11} To perform
ensemble refinement ``on the fly'', or to enhance the sampling of
relevant configurations in cases where the reference distribution and
the optimal ensemble density have limited overlap, we considered
replica-simulation methods in which a restraint is imposed through a
biasing potential that acts on averages over all replicas.  We showed
using a mean-field treatment that to obtain a size-consistent result,
the biasing potential has to be scaled by the number $N$ of replicas,
i.e., the restraint has to become stiffer as more replicas are
included. Then, the Bayesian replica ensemble refinement converges to
the optimal Bayesian ensemble refinement in the limit of infinitely
many replicas, $N\rightarrow\infty$.  This result clarifies the need
to scale the biasing potential, which arises also in maximum entropy
treatments with strict
constraints,\cite{Cavalli:13:1,Cavalli:Erratum:13,RouxWeare:13} with
the number of replicas $N$ to obtain a size-consistent result. An
adaptive method, as described in the Appendix, provides a possible
alternative to replica-based approaches.

The BioEn approach combines the replica and EROS refinement
methods. The replica simulations are used to create an enriched sample
of configurations.  A free-energy calculation is used to determine the
appropriate weights according to the reference ensemble. The optimal
weights according to Bayesian ensemble refinement are then determined
by EROS.  This combined approach addresses the shortcomings of either
method, i.e., the need to work with relatively small $N$ in replica
simulations, and potentially limited overlap of reference and
optimized distribution in EROS.  Using free-energy reweighting
methods, it may also be possible to include configurations from other
types of ensemble-biased simulations, including those designed to
satisfy measurements
exactly.\cite{Camilloni:JCTC:13,Hansen:14,Marinelli:15} Because of the
flexibility and expected rapid convergence, the BioEn method combining
Bayesian replica and EROS refinement should perform well in practical
applications.

In two examples that are analytically tractable and one requiring
numerical calculations, we demonstrated the equivalence
of the different methods in the appropriate limits.  We also studied
the convergence properties of Bayesian replica simulations with the
number of replicas $N$, and of EROS reweighting with the sample size
$n$.  Our examples showed similar convergence of the log-probability
of the two refinement approaches to the optimal limit as $1/N$ and
$1/n$, respectively.

The BioEn approach also addresses a major issue in Bayesian
ensemble refinement, namely the choice of $\theta$.  This parameter
enters the prior to express our confidence in the reference
distribution.  We find that in the optimal Bayesian ensemble
distributions, a change in $\theta$ is simply equivalent to a uniform
scaling of all squared Gaussian errors $\sigma_i^2$.  Since EROS reweighting is
usually orders of magnitudes less costly than sampling multiple
replicas in coupled molecular simulations, one can efficiently obtain
estimates of the relative entropy $S_\mathrm{KL}$ for different
$\theta$.  From plots of $S_\mathrm{KL}$ against $\chi^2$ one can make
an educated choice of $\theta$, as in other regularization approaches
to inverse problems.\cite{Hansen:93}

Finally, the reweighting of individual structures, either directly
using EROS or in the combined approach, should prove useful in the
optimization of potential energy functions by fitting them to
experimental data (see, e.g.,
Refs. \onlinecite{Norgaard:08,Li:11,Wang:13}).  If one has a good
understanding of the sources of the errors in the energy surface
$U(\X)$, parameters in $U$ can be fitted directly, as was done, e.g.,
for the star force fields of proteins \cite{Best:09} and for
RNA.\cite{Chen:13} At the other extreme, Bayesian approaches have been
used before to infer entire energy functions.\cite{Habeck:14} Here, we
suggest to concentrate on the change in weight of structures
$\X_\alpha$, i.e., $\ln \Wopt/w_\alpha^0=-\beta \Delta U_\alpha +
\mathit{const.}$, which defines the required change $\Delta U_\alpha$
in the potential energy to match experiment.  By examining the
correlation of this force field error $\Delta U_\alpha$ with elements
of the force field (e.g., peptide dihedral angles \cite{Best:09} or
base stacking interactions \cite{Chen:13}), it might be possible to
identify sources of the error and then correct for them.

It is important to emphasize that a number of assumptions enter the
ensemble refinement procedure.  The central (and declared!) assumption
is that of a reference distribution.  Here it may be possible to use
combinations of multiple potential energy functions $u_k(\X)$,
representing different force fields or conditions $k$, that jointly
cover the relevant phase space better than any potential alone.  One
way to mix such potentials is by using a multistate
model,\cite{best-2005-4} $U(\X) = -\gamma^{-1}\ln \sum_k \exp[-\gamma
u_k(\X)+\epsilon_k]$, where $\gamma$ is the mixing ``temperature'' and
$\epsilon_k$ are energy offsets that weight the different force
fields.  Another important challenge is that one has to estimate
errors both in the measurements and in the calculation of the
observables. Procedures to account for uncertainties in the error
estimates have been developed.\cite{Rieping:05} Within the present
framework, one could include error distributions in the maximization
of the log-posterior, or average over optimal solutions obtained for
different errors.  In addition, in many cases the Gaussian error model
may not be appropriate. As discussed, to handle more general error
models, one can substitute the log-likelihood $\ln {\cal
  P}[\mathrm{data}|p(\X)]$ for $-\chi^2/2$ in ${\cal
  P}[p(\X)|\mathrm{data}]$.

Overall, we expect our exploration of different Bayesian ensemble
refinement approaches to serve both as a basis for practical
applications and as a starting point for further investigations.  In
particular, we have here not considered an orthogonal refinement
approach in which one seeks to represent the ensemble by a minimal set
of structures.\cite{Boura:11,Berlin:13,Cossio:13,Pelikan:09}  As we
had shown before, EROS and minimal ensemble refinement, properly
interpreted, can give consistent results.\cite{Francis:11}  However,
the relation of the different methods is not well understood, e.g.,
concerning the limiting behavior for large sample sizes.

\acknowledgments We thank Drs. Pilar Cossio and Roberto Covino, and
Profs. Andrea Cavalli, Kresten Lindorff-Larsen, Beno\^{i}t Roux, Andrej Sali, and
Michele Vendruscolo for helpful discussions. This work was supported
by the Max Planck Society.

\appendix
\section{Adaptive sampling of optimal Bayesian distribution without replicas}
\label{app:A}

We define probability densities of the observables alone by integrating out all other
degrees of freedom,
\begin{eqnarray}
  \label{eq:A1}
  P_0(\Y) &=&\int d\X\,p_0(\X)\prod_{i=1}^M\delta[y_i-y_i(\X)],\\
  P(\Y)&=&\int d\X\,p(\X)\prod_{i=1}^M\delta[y_i-y_i(\X)]. 
\end{eqnarray}
According to Eqs.~(\ref{eq:34}) and (\ref{eq:34b}), these two
distributions are related to each other,
\begin{equation}
  \label{eq:A2}
  P(\Y)\propto P_0(\Y)
  \exp{\left(-\frac{1}{\theta}\sum_{i,j=1}^M y_i
      (\bm{\Sigma}^{-1})_{ij}f_j\right)}
\end{equation}
for possibly correlated Gaussian errors, where the generalized forces
$f_j$ have to be determined self-consistently such that
\begin{equation}
  \label{eq:A3}
  f_j=\int d\Y\,P(\Y)y_j-Y_j^{\OBS} = \langle y_j\rangle-Y_j^{\OBS}.
\end{equation}
As a consequence, $p(\X)\propto
p_0(\X)\exp[-\theta^{-1}\Y^T(\X)\bm{\Sigma}^{-1}\mathbf{f}]$ where
superscript $T$ indicates the transpose in vector-matrix notation.
The biasing potential thus assumes a functional form linear in the
$y_i(\X)$, as seen in standard maximum entropy approaches (see, e.g.,
Refs. \onlinecite{Pitera:12,Beauchamp:14,White:14}), but the
generalized forces $f_i$ take on different values here.  We also note
that the forces $f_j$ defining the optimal distribution can be
interpreted mechanically.  With Eq.~(\ref{eq:A2}) one finds that the
mean force trying to ``restore'' the reference distribution,
$\mathbf{F}^{(\mathrm{ref})} \equiv \int d\Y\,P(\Y) \partial [-\ln
P_0(\Y)] /\partial\Y=-\theta^{-1} \bm{\Sigma}^{-1}\mathbf{f}$, is
exactly balanced by the mean force to fit the data,
$\mathbf{F}^{(\mathrm{fit})} \equiv \int d\Y\,P(\Y) \partial
(\chi^2/2) /\partial\Y= \bm{\Sigma}^{-1}\mathbf{f}$, up to a factor
$\theta$, with $\chi^2$ from Eq.~(\ref{eq:4}) with $Y_j^{\OBS}$
instead of $y_j^{\OBS}$.

Formally, the $f_j$ can be obtained by solving $M$ coupled nonlinear
equations.  We define generating functions
$\phi_0(\Z)\equiv\int d\Y\,P_0(\Y)\exp(\Y\cdot\Z)=\int d\X\,
p_0(\X)\exp[\sum_i y_i(\X) z_i]$
and
$\phi(\Z)\equiv\int d\Y\,P(\Y)\exp(\Y\cdot\Z)=\int d\X\,
p(\X)\exp[\sum_i y_i(\X) z_i]$,
assuming that the integrals exist.  Multiplying Eq.~(\ref{eq:A2}) by
$\exp(\Y\cdot\Z)$ and integrating over $\Y$, we obtain
\begin{equation}
  \label{eq:A4}
  \phi(\Z)=\frac{\phi_0\left[\Z-\theta^{-1}(\bm{\Sigma}^{-1})\F\right]}
  {\phi_0\left[-\theta^{-1}(\bm{\Sigma}^{-1})\F\right]},
\end{equation}
where the denominator ensures normalization, $\phi(0)=1$. With
$\langle
y_j\rangle = \left.\partial\phi(\Z)/{\partial z_j}\right|_{\Z=0}$,
Eq.~(\ref{eq:A3}) for the vector of forces $\F$ becomes
satisfy
\begin{equation}
  \label{eq:A5}
  \F =\left.\frac{\partial\ln\phi_0\left[\Z-\theta^{-1}(\bm{\Sigma}^{-1})\F\right]}{\partial \mathbf{z}}\right|_{\Z=0}-\mathbf{Y}^{\OBS},
\end{equation}
where $\ln \phi_0(\Z)$ is the cumulant generating function of the
reference distribution of observables.

In cases where the equations cannot be solved directly, one can
determine the force-vector $\F$ adaptively. In the following, we
present a simple algorithm that can be combined with existing
simulation procedures. This approach is related to that of White and
Voth,\cite{White:14} in which an adaptive gradient-based method is
used to construct a distribution in which the $y_j$-averages exactly
match $Y_j^{\OBS}$. Here, by contrast, we include measurement errors
and thus do not demand exact agreement with the observed values.
Instead, the $f_j$ have to be determined self-consistently to satisfy
Eq. (\ref{eq:A3}). In our adaptive optimization, we adjust the
generalized forces $f_j$ ``on the fly'' according to the running
averages of the $y_j$,
\begin{equation}
  \label{eq:A6}
  \F(t)=\frac{1}{t}\int_0^t d\tau\,\Y[\X(\tau)] - \mathbf{Y}^{\OBS},
\end{equation}
with initial value $\F(0)=\Y[\X(0)]$. Here, the trajectory $\X(t)$
evolves according to the time-dependent potential energy
$U(\X)+\Y(\X) \bm{\Sigma}^{-1} \F(t)/\theta$.  We note that by extending the
phase space to include both $\X$ and $\F$, this algorithm can be cast
in a Markovian form.  If $L_f$ is the Liouville evolution operator for
the phase space density of $\X$ according to the
molecular dynamics or Monte Carlo simulation protocol and potential
$U(\X)+\Y(\X) \bm{\Sigma}^{-1} \F/\theta$ with fixed $\F$, then
the extended phase space density $\rho=\rho(\X,\F,t)$ satisfies a
Markovian Liouville-type evolution equation
\begin{equation}
  \label{eq:A7}
  \frac{\partial \rho}{\partial t} =
  \left[ L_f +
    \frac{1}{t}\frac{\partial}{\partial\F}\left(\F+\mathbf{Y}^{\OBS}-\Y(\X)\right)\right] 
  \rho.
\end{equation}
Using this relation, one can show that for the Gaussian example in the main text,
with the mean as observable and overdamped diffusion for the dynamics of $\X$, the
adaptive sampling is globally converging to the optimal Bayesian
ensemble distribution. For the example in Fig.~\ref{fig:2}, with Monte
Carlo sampling of $x$, we observed convergence numerically.

We note that in the adaptive determination of the generalized forces
$f_j$ defining the posterior distribution, variants of
Eq.~(\ref{eq:A6}) are possible.  In particular, one can average
between an initial guess $\F_0$ and the evolving mean, e.g., as
$\F(t)=w(t)\F_0+[1-w(t)][t^{-1}\int_0^t d\tau\,\Y[\X(\tau)] -
\mathbf{Y}^{\OBS}]$, where $w(t)$ is a weight function that decreases
to zero with time, e.g., $w(t)=\exp(-t/t_0)$ for a suitably chosen
relaxation time $t_0$.


\begin{thebibliography}{55}%
\makeatletter
\providecommand \@ifxundefined [1]{%
 \@ifx{#1\undefined}
}%
\providecommand \@ifnum [1]{%
 \ifnum #1\expandafter \@firstoftwo
 \else \expandafter \@secondoftwo
 \fi
}%
\providecommand \@ifx [1]{%
 \ifx #1\expandafter \@firstoftwo
 \else \expandafter \@secondoftwo
 \fi
}%
\providecommand \natexlab [1]{#1}%
\providecommand \enquote  [1]{``#1''}%
\providecommand \bibnamefont  [1]{#1}%
\providecommand \bibfnamefont [1]{#1}%
\providecommand \citenamefont [1]{#1}%
\providecommand \href@noop [0]{\@secondoftwo}%
\providecommand \href [0]{\begingroup \@sanitize@url \@href}%
\providecommand \@href[1]{\@@startlink{#1}\@@href}%
\providecommand \@@href[1]{\endgroup#1\@@endlink}%
\providecommand \@sanitize@url [0]{\catcode `\\12\catcode `\$12\catcode
  `\&12\catcode `\#12\catcode `\^12\catcode `\_12\catcode `\%12\relax}%
\providecommand \@@startlink[1]{}%
\providecommand \@@endlink[0]{}%
\providecommand \url  [0]{\begingroup\@sanitize@url \@url }%
\providecommand \@url [1]{\endgroup\@href {#1}{\urlprefix }}%
\providecommand \urlprefix  [0]{URL }%
\providecommand \Eprint [0]{\href }%
\providecommand \doibase [0]{http://dx.doi.org/}%
\providecommand \selectlanguage [0]{\@gobble}%
\providecommand \bibinfo  [0]{\@secondoftwo}%
\providecommand \bibfield  [0]{\@secondoftwo}%
\providecommand \translation [1]{[#1]}%
\providecommand \BibitemOpen [0]{}%
\providecommand \bibitemStop [0]{}%
\providecommand \bibitemNoStop [0]{.\EOS\space}%
\providecommand \EOS [0]{\spacefactor3000\relax}%
\providecommand \BibitemShut  [1]{\csname bibitem#1\endcsname}%
\let\auto@bib@innerbib\@empty
\bibitem [{\citenamefont {Boomsma}, \citenamefont {Ferkinghoff-Borg},\ and\
  \citenamefont {Lindorff-Larsen}(2014)}]{Boomsma:14:1}%
  \BibitemOpen
  \bibfield  {author} {\bibinfo {author} {\bibfnamefont {W.}~\bibnamefont
  {Boomsma}}, \bibinfo {author} {\bibfnamefont {J.}~\bibnamefont
  {Ferkinghoff-Borg}}, \ and\ \bibinfo {author} {\bibfnamefont
  {K.}~\bibnamefont {Lindorff-Larsen}},\ }\href@noop {} {\bibfield  {journal}
  {\bibinfo  {journal} {PLoS Comp. Biology}\ }\textbf {\bibinfo {volume} {10}}
  (\bibinfo {year} {2014})}\BibitemShut {NoStop}%
\bibitem [{\citenamefont {Sali}\ \emph {et~al.}(2015)\citenamefont {Sali},
  \citenamefont {Berman}, \citenamefont {Schwede}, \citenamefont {Trewhella},
  \citenamefont {Kleywegt}, \citenamefont {Burley}, \citenamefont {Markley},
  \citenamefont {Nakamura}, \citenamefont {Adams}, \citenamefont {Bonvin},
  \citenamefont {Chiu}, \citenamefont {Peraro}, \citenamefont {Di~Maio},
  \citenamefont {Ferrin}, \citenamefont {Gr{\"u}newald}, \citenamefont
  {Gutmanas}, \citenamefont {Henderson}, \citenamefont {Hummer}, \citenamefont
  {Iwasaki}, \citenamefont {Johnson}, \citenamefont {Lawson}, \citenamefont
  {Meiler}, \citenamefont {Marti-Renom}, \citenamefont {Montelione},
  \citenamefont {Nilges}, \citenamefont {Nussinov}, \citenamefont {Patwardhan},
  \citenamefont {Rappsilber}, \citenamefont {Read}, \citenamefont {Saibil},
  \citenamefont {Schr{\"o}der}, \citenamefont {Schwieters}, \citenamefont
  {Seidel}, \citenamefont {Svergun}, \citenamefont {Topf}, \citenamefont
  {Ulrich}, \citenamefont {Velankar},\ and\ \citenamefont
  {Westbrook}}]{wwPDB:Structure:2015}%
  \BibitemOpen
  \bibfield  {author} {\bibinfo {author} {\bibfnamefont {A.}~\bibnamefont
  {Sali}}, \bibinfo {author} {\bibfnamefont {H.~M.}\ \bibnamefont {Berman}},
  \bibinfo {author} {\bibfnamefont {T.}~\bibnamefont {Schwede}}, \bibinfo
  {author} {\bibfnamefont {J.}~\bibnamefont {Trewhella}}, \bibinfo {author}
  {\bibfnamefont {G.}~\bibnamefont {Kleywegt}}, \bibinfo {author}
  {\bibfnamefont {S.~K.}\ \bibnamefont {Burley}}, \bibinfo {author}
  {\bibfnamefont {J.}~\bibnamefont {Markley}}, \bibinfo {author} {\bibfnamefont
  {H.}~\bibnamefont {Nakamura}}, \bibinfo {author} {\bibfnamefont
  {P.}~\bibnamefont {Adams}}, \bibinfo {author} {\bibfnamefont {A.~M. J.~J.}\
  \bibnamefont {Bonvin}}, \bibinfo {author} {\bibfnamefont {W.}~\bibnamefont
  {Chiu}}, \bibinfo {author} {\bibfnamefont {M.}~\bibnamefont {Peraro}},
  \bibinfo {author} {\bibfnamefont {F.}~\bibnamefont {Di~Maio}}, \bibinfo
  {author} {\bibfnamefont {T.~E.}\ \bibnamefont {Ferrin}}, \bibinfo {author}
  {\bibfnamefont {K.}~\bibnamefont {Gr{\"u}newald}}, \bibinfo {author}
  {\bibfnamefont {A.}~\bibnamefont {Gutmanas}}, \bibinfo {author}
  {\bibfnamefont {R.}~\bibnamefont {Henderson}}, \bibinfo {author}
  {\bibfnamefont {G.}~\bibnamefont {Hummer}}, \bibinfo {author} {\bibfnamefont
  {K.}~\bibnamefont {Iwasaki}}, \bibinfo {author} {\bibfnamefont
  {G.}~\bibnamefont {Johnson}}, \bibinfo {author} {\bibfnamefont
  {C.}~\bibnamefont {Lawson}}, \bibinfo {author} {\bibfnamefont
  {J.}~\bibnamefont {Meiler}}, \bibinfo {author} {\bibfnamefont {M.~A.}\
  \bibnamefont {Marti-Renom}}, \bibinfo {author} {\bibfnamefont
  {G.}~\bibnamefont {Montelione}}, \bibinfo {author} {\bibfnamefont
  {M.}~\bibnamefont {Nilges}}, \bibinfo {author} {\bibfnamefont
  {R.}~\bibnamefont {Nussinov}}, \bibinfo {author} {\bibfnamefont
  {A.}~\bibnamefont {Patwardhan}}, \bibinfo {author} {\bibfnamefont
  {J.}~\bibnamefont {Rappsilber}}, \bibinfo {author} {\bibfnamefont {R.~J.}\
  \bibnamefont {Read}}, \bibinfo {author} {\bibfnamefont {H.}~\bibnamefont
  {Saibil}}, \bibinfo {author} {\bibfnamefont {G.~F.}\ \bibnamefont
  {Schr{\"o}der}}, \bibinfo {author} {\bibfnamefont {C.~D.}\ \bibnamefont
  {Schwieters}}, \bibinfo {author} {\bibfnamefont {C.~A.~M.}\ \bibnamefont
  {Seidel}}, \bibinfo {author} {\bibfnamefont {D.}~\bibnamefont {Svergun}},
  \bibinfo {author} {\bibfnamefont {M.}~\bibnamefont {Topf}}, \bibinfo {author}
  {\bibfnamefont {E.~L.}\ \bibnamefont {Ulrich}}, \bibinfo {author}
  {\bibfnamefont {S.}~\bibnamefont {Velankar}}, \ and\ \bibinfo {author}
  {\bibfnamefont {J.~D.}\ \bibnamefont {Westbrook}},\ }\href {\doibase
  10.1016/j.str.2015.05.013} {\bibfield  {journal} {\bibinfo  {journal}
  {Structure}\ }\textbf {\bibinfo {volume} {23}},\ \bibinfo {pages} {1156}
  (\bibinfo {year} {2015})}\BibitemShut {NoStop}%
\bibitem [{\citenamefont {Boura}\ \emph {et~al.}(2011)\citenamefont {Boura},
  \citenamefont {Rozycki}, \citenamefont {Herrick}, \citenamefont {Chung},
  \citenamefont {Vecer}, \citenamefont {Eaton}, \citenamefont {Cafiso},
  \citenamefont {Hummer},\ and\ \citenamefont {Hurley}}]{Boura:11}%
  \BibitemOpen
  \bibfield  {author} {\bibinfo {author} {\bibfnamefont {E.}~\bibnamefont
  {Boura}}, \bibinfo {author} {\bibfnamefont {B.}~\bibnamefont {Rozycki}},
  \bibinfo {author} {\bibfnamefont {D.~Z.}\ \bibnamefont {Herrick}}, \bibinfo
  {author} {\bibfnamefont {H.~S.}\ \bibnamefont {Chung}}, \bibinfo {author}
  {\bibfnamefont {J.}~\bibnamefont {Vecer}}, \bibinfo {author} {\bibfnamefont
  {W.~A.}\ \bibnamefont {Eaton}}, \bibinfo {author} {\bibfnamefont {D.~S.}\
  \bibnamefont {Cafiso}}, \bibinfo {author} {\bibfnamefont {G.}~\bibnamefont
  {Hummer}}, \ and\ \bibinfo {author} {\bibfnamefont {J.~H.}\ \bibnamefont
  {Hurley}},\ }\href@noop {} {\bibfield  {journal} {\bibinfo  {journal} {Proc.
  Natl. Acad. Sci. U.S.A.}\ }\textbf {\bibinfo {volume} {108}},\ \bibinfo
  {pages} {9437} (\bibinfo {year} {2011})}\BibitemShut {NoStop}%
\bibitem [{\citenamefont {Ward}, \citenamefont {Sali},\ and\ \citenamefont
  {Wilson}(2013)}]{Sali:Science:13}%
  \BibitemOpen
  \bibfield  {author} {\bibinfo {author} {\bibfnamefont {A.~B.}\ \bibnamefont
  {Ward}}, \bibinfo {author} {\bibfnamefont {A.}~\bibnamefont {Sali}}, \ and\
  \bibinfo {author} {\bibfnamefont {I.~A.}\ \bibnamefont {Wilson}},\ }\href
  {\doibase 10.1126/science.1228565} {\bibfield  {journal} {\bibinfo  {journal}
  {Science}\ }\textbf {\bibinfo {volume} {339}},\ \bibinfo {pages} {913}
  (\bibinfo {year} {2013})}\BibitemShut {NoStop}%
\bibitem [{\citenamefont {Camilloni}\ and\ \citenamefont
  {Vendruscolo}(2014)}]{Camilloni:14}%
  \BibitemOpen
  \bibfield  {author} {\bibinfo {author} {\bibfnamefont {C.}~\bibnamefont
  {Camilloni}}\ and\ \bibinfo {author} {\bibfnamefont {M.}~\bibnamefont
  {Vendruscolo}},\ }\href@noop {} {\bibfield  {journal} {\bibinfo  {journal}
  {J. Am. Chem. Soc.}\ }\textbf {\bibinfo {volume} {136}},\ \bibinfo {pages}
  {8982} (\bibinfo {year} {2014})}\BibitemShut {NoStop}%
\bibitem [{\citenamefont {Fisher}, \citenamefont {Huang},\ and\ \citenamefont
  {Stultz}(2010)}]{Fisher:10}%
  \BibitemOpen
  \bibfield  {author} {\bibinfo {author} {\bibfnamefont {C.~K.}\ \bibnamefont
  {Fisher}}, \bibinfo {author} {\bibfnamefont {A.}~\bibnamefont {Huang}}, \
  and\ \bibinfo {author} {\bibfnamefont {C.~M.}\ \bibnamefont {Stultz}},\
  }\href@noop {} {\bibfield  {journal} {\bibinfo  {journal} {J. Am. Chem.
  Soc.}\ }\textbf {\bibinfo {volume} {132}},\ \bibinfo {pages} {14919}
  (\bibinfo {year} {2010})}\BibitemShut {NoStop}%
\bibitem [{\citenamefont {Fisher}\ and\ \citenamefont
  {Stultz}(2011)}]{Fisher:11}%
  \BibitemOpen
  \bibfield  {author} {\bibinfo {author} {\bibfnamefont {C.~K.}\ \bibnamefont
  {Fisher}}\ and\ \bibinfo {author} {\bibfnamefont {C.~M.}\ \bibnamefont
  {Stultz}},\ }\href@noop {} {\bibfield  {journal} {\bibinfo  {journal} {Curr.
  Opin. Struct. Biol.}\ }\textbf {\bibinfo {volume} {21}},\ \bibinfo {pages}
  {426} (\bibinfo {year} {2011})}\BibitemShut {NoStop}%
\bibitem [{\citenamefont {Sanchez-Martinez}\ and\ \citenamefont
  {Crehuet}(2014)}]{Sanchez-Martinez:14}%
  \BibitemOpen
  \bibfield  {author} {\bibinfo {author} {\bibfnamefont {M.}~\bibnamefont
  {Sanchez-Martinez}}\ and\ \bibinfo {author} {\bibfnamefont {R.}~\bibnamefont
  {Crehuet}},\ }\href@noop {} {\bibfield  {journal} {\bibinfo  {journal} {Phys.
  Chem. Chem. Phys.}\ }\textbf {\bibinfo {volume} {16}},\ \bibinfo {pages}
  {26030} (\bibinfo {year} {2014})}\BibitemShut {NoStop}%
\bibitem [{\citenamefont {Dedmon}\ \emph {et~al.}(2005)\citenamefont {Dedmon},
  \citenamefont {Lindorff-Larsen}, \citenamefont {Christodoulou}, \citenamefont
  {Vendruscolo},\ and\ \citenamefont {Dobson}}]{Dedmon:05}%
  \BibitemOpen
  \bibfield  {author} {\bibinfo {author} {\bibfnamefont {M.~M.}\ \bibnamefont
  {Dedmon}}, \bibinfo {author} {\bibfnamefont {K.}~\bibnamefont
  {Lindorff-Larsen}}, \bibinfo {author} {\bibfnamefont {J.}~\bibnamefont
  {Christodoulou}}, \bibinfo {author} {\bibfnamefont {M.}~\bibnamefont
  {Vendruscolo}}, \ and\ \bibinfo {author} {\bibfnamefont {C.~M.}\ \bibnamefont
  {Dobson}},\ }\href@noop {} {\bibfield  {journal} {\bibinfo  {journal} {J. Am.
  Chem. Soc.}\ }\textbf {\bibinfo {volume} {127}},\ \bibinfo {pages} {476}
  (\bibinfo {year} {2005})}\BibitemShut {NoStop}%
\bibitem [{\citenamefont {Mantsyzov}\ \emph {et~al.}(2014)\citenamefont
  {Mantsyzov}, \citenamefont {Maltsev}, \citenamefont {Ying}, \citenamefont
  {Shen}, \citenamefont {Hummer},\ and\ \citenamefont {Bax}}]{Mantsyzov:14}%
  \BibitemOpen
  \bibfield  {author} {\bibinfo {author} {\bibfnamefont {A.~B.}\ \bibnamefont
  {Mantsyzov}}, \bibinfo {author} {\bibfnamefont {A.~S.}\ \bibnamefont
  {Maltsev}}, \bibinfo {author} {\bibfnamefont {J.}~\bibnamefont {Ying}},
  \bibinfo {author} {\bibfnamefont {Y.}~\bibnamefont {Shen}}, \bibinfo {author}
  {\bibfnamefont {G.}~\bibnamefont {Hummer}}, \ and\ \bibinfo {author}
  {\bibfnamefont {A.}~\bibnamefont {Bax}},\ }\href {\doibase 10.1002/pro.2511}
  {\bibfield  {journal} {\bibinfo  {journal} {Protein Sci.}\ }\textbf {\bibinfo
  {volume} {23}},\ \bibinfo {pages} {1275} (\bibinfo {year} {2014 2014
  2014})}\BibitemShut {NoStop}%
\bibitem [{\citenamefont {Mantsyzov}\ \emph {et~al.}(2015)\citenamefont
  {Mantsyzov}, \citenamefont {Shen}, \citenamefont {Lee}, \citenamefont
  {Hummer},\ and\ \citenamefont {Bax}}]{Mantsyzov:15}%
  \BibitemOpen
  \bibfield  {author} {\bibinfo {author} {\bibfnamefont {A.~B.}\ \bibnamefont
  {Mantsyzov}}, \bibinfo {author} {\bibfnamefont {Y.}~\bibnamefont {Shen}},
  \bibinfo {author} {\bibfnamefont {J.}~\bibnamefont {Lee}}, \bibinfo {author}
  {\bibfnamefont {G.}~\bibnamefont {Hummer}}, \ and\ \bibinfo {author}
  {\bibfnamefont {A.}~\bibnamefont {Bax}},\ }\href {\doibase
  10.1007/s10858-015-9971-2} {\bibfield  {journal} {\bibinfo  {journal} {J.
  Biomol. NMR}\ ,\ \bibinfo {pages} {1}} (\bibinfo {year} {2015})}\BibitemShut
  {NoStop}%
\bibitem [{\citenamefont {Schr\"{o}der}(2015)}]{Schroder:15}%
  \BibitemOpen
  \bibfield  {author} {\bibinfo {author} {\bibfnamefont {G.~F.}\ \bibnamefont
  {Schr\"{o}der}},\ }\href@noop {} {\bibfield  {journal} {\bibinfo  {journal}
  {Curr. Opin. Struct. Biol.}\ }\textbf {\bibinfo {volume} {31}},\ \bibinfo
  {pages} {20} (\bibinfo {year} {2015})}\BibitemShut {NoStop}%
\bibitem [{\citenamefont {Rozycki}, \citenamefont {Kim},\ and\ \citenamefont
  {Hummer}(2011)}]{Rozycki:11}%
  \BibitemOpen
  \bibfield  {author} {\bibinfo {author} {\bibfnamefont {B.}~\bibnamefont
  {Rozycki}}, \bibinfo {author} {\bibfnamefont {Y.~C.}\ \bibnamefont {Kim}}, \
  and\ \bibinfo {author} {\bibfnamefont {G.}~\bibnamefont {Hummer}},\
  }\href@noop {} {\bibfield  {journal} {\bibinfo  {journal} {Structure}\
  }\textbf {\bibinfo {volume} {19}},\ \bibinfo {pages} {109} (\bibinfo {year}
  {2011})}\BibitemShut {NoStop}%
\bibitem [{\citenamefont {Francis}\ \emph {et~al.}(2011)\citenamefont
  {Francis}, \citenamefont {Rozycki}, \citenamefont {Koveal}, \citenamefont
  {Hummer}, \citenamefont {Page},\ and\ \citenamefont {Peti}}]{Francis:11}%
  \BibitemOpen
  \bibfield  {author} {\bibinfo {author} {\bibfnamefont {D.~M.}\ \bibnamefont
  {Francis}}, \bibinfo {author} {\bibfnamefont {B.}~\bibnamefont {Rozycki}},
  \bibinfo {author} {\bibfnamefont {D.}~\bibnamefont {Koveal}}, \bibinfo
  {author} {\bibfnamefont {G.}~\bibnamefont {Hummer}}, \bibinfo {author}
  {\bibfnamefont {R.}~\bibnamefont {Page}}, \ and\ \bibinfo {author}
  {\bibfnamefont {W.}~\bibnamefont {Peti}},\ }\href@noop {} {\bibfield
  {journal} {\bibinfo  {journal} {Nature Chem. Biology}\ }\textbf {\bibinfo
  {volume} {7}},\ \bibinfo {pages} {916} (\bibinfo {year} {2011})}\BibitemShut
  {NoStop}%
\bibitem [{\citenamefont {Scheek}\ \emph {et~al.}(1991)\citenamefont {Scheek},
  \citenamefont {Torda}, \citenamefont {Kemmink},\ and\ \citenamefont {{van
  Gunsteren}}}]{Scheek:91}%
  \BibitemOpen
  \bibfield  {author} {\bibinfo {author} {\bibfnamefont {R.~M.}\ \bibnamefont
  {Scheek}}, \bibinfo {author} {\bibfnamefont {A.~E.}\ \bibnamefont {Torda}},
  \bibinfo {author} {\bibfnamefont {J.}~\bibnamefont {Kemmink}}, \ and\
  \bibinfo {author} {\bibfnamefont {W.~F.}\ \bibnamefont {{van Gunsteren}}},\
  }\href@noop {} {\bibfield  {journal} {\bibinfo  {journal} {{NATO} Advanced
  Science Institutes Series Series A Life Sciences}\ }\textbf {\bibinfo
  {volume} {225}},\ \bibinfo {pages} {209} (\bibinfo {year}
  {1991})}\BibitemShut {NoStop}%
\bibitem [{\citenamefont {Lange}\ \emph {et~al.}(2008)\citenamefont {Lange},
  \citenamefont {Lakomek}, \citenamefont {Fares}, \citenamefont {Schr\"{o}der},
  \citenamefont {Walter}, \citenamefont {Becker}, \citenamefont {Meiler},
  \citenamefont {Grubm\"{u}ller}, \citenamefont {Griesinger},\ and\
  \citenamefont {{de Groot}}}]{Lange:08}%
  \BibitemOpen
  \bibfield  {author} {\bibinfo {author} {\bibfnamefont {O.~F.}\ \bibnamefont
  {Lange}}, \bibinfo {author} {\bibfnamefont {N.~A.}\ \bibnamefont {Lakomek}},
  \bibinfo {author} {\bibfnamefont {C.}~\bibnamefont {Fares}}, \bibinfo
  {author} {\bibfnamefont {G.~F.}\ \bibnamefont {Schr\"{o}der}}, \bibinfo
  {author} {\bibfnamefont {K.~F.~A.}\ \bibnamefont {Walter}}, \bibinfo {author}
  {\bibfnamefont {S.}~\bibnamefont {Becker}}, \bibinfo {author} {\bibfnamefont
  {J.}~\bibnamefont {Meiler}}, \bibinfo {author} {\bibfnamefont
  {H.}~\bibnamefont {Grubm\"{u}ller}}, \bibinfo {author} {\bibfnamefont
  {C.}~\bibnamefont {Griesinger}}, \ and\ \bibinfo {author} {\bibfnamefont
  {B.~L.}\ \bibnamefont {{de Groot}}},\ }\href@noop {} {\bibfield  {journal}
  {\bibinfo  {journal} {Science}\ }\textbf {\bibinfo {volume} {320}},\ \bibinfo
  {pages} {1471} (\bibinfo {year} {2008})}\BibitemShut {NoStop}%
\bibitem [{\citenamefont {Olsson}\ \emph {et~al.}(2014)\citenamefont {Olsson},
  \citenamefont {V\"{o}geli}, \citenamefont {Cavalli}, \citenamefont {Boomsma},
  \citenamefont {Ferkinghoff-Borg}, \citenamefont {Lindorff-Larsen},\ and\
  \citenamefont {Hamelryck}}]{Olsson:14}%
  \BibitemOpen
  \bibfield  {author} {\bibinfo {author} {\bibfnamefont {S.}~\bibnamefont
  {Olsson}}, \bibinfo {author} {\bibfnamefont {B.~R.}\ \bibnamefont
  {V\"{o}geli}}, \bibinfo {author} {\bibfnamefont {A.}~\bibnamefont {Cavalli}},
  \bibinfo {author} {\bibfnamefont {W.}~\bibnamefont {Boomsma}}, \bibinfo
  {author} {\bibfnamefont {J.}~\bibnamefont {Ferkinghoff-Borg}}, \bibinfo
  {author} {\bibfnamefont {K.}~\bibnamefont {Lindorff-Larsen}}, \ and\ \bibinfo
  {author} {\bibfnamefont {T.}~\bibnamefont {Hamelryck}},\ }\href@noop {}
  {\bibfield  {journal} {\bibinfo  {journal} {J. Chem. Theory Comput.}\
  }\textbf {\bibinfo {volume} {10}},\ \bibinfo {pages} {3484} (\bibinfo {year}
  {2014})}\BibitemShut {NoStop}%
\bibitem [{\citenamefont {Boura}\ \emph {et~al.}(2012)\citenamefont {Boura},
  \citenamefont {Rozycki}, \citenamefont {Chung}, \citenamefont {Herrick},
  \citenamefont {Canagarajah}, \citenamefont {Cafiso}, \citenamefont {Eaton},
  \citenamefont {Hummer},\ and\ \citenamefont {Hurley}}]{Boura:12}%
  \BibitemOpen
  \bibfield  {author} {\bibinfo {author} {\bibfnamefont {E.}~\bibnamefont
  {Boura}}, \bibinfo {author} {\bibfnamefont {B.}~\bibnamefont {Rozycki}},
  \bibinfo {author} {\bibfnamefont {H.~S.}\ \bibnamefont {Chung}}, \bibinfo
  {author} {\bibfnamefont {D.~Z.}\ \bibnamefont {Herrick}}, \bibinfo {author}
  {\bibfnamefont {B.}~\bibnamefont {Canagarajah}}, \bibinfo {author}
  {\bibfnamefont {D.~S.}\ \bibnamefont {Cafiso}}, \bibinfo {author}
  {\bibfnamefont {W.~A.}\ \bibnamefont {Eaton}}, \bibinfo {author}
  {\bibfnamefont {G.}~\bibnamefont {Hummer}}, \ and\ \bibinfo {author}
  {\bibfnamefont {J.~H.}\ \bibnamefont {Hurley}},\ }\href@noop {} {\bibfield
  {journal} {\bibinfo  {journal} {Structure}\ }\textbf {\bibinfo {volume}
  {20}},\ \bibinfo {pages} {874} (\bibinfo {year} {2012})}\BibitemShut
  {NoStop}%
\bibitem [{\citenamefont {Cossio}\ and\ \citenamefont
  {Hummer}(2013)}]{Cossio:13}%
  \BibitemOpen
  \bibfield  {author} {\bibinfo {author} {\bibfnamefont {P.}~\bibnamefont
  {Cossio}}\ and\ \bibinfo {author} {\bibfnamefont {G.}~\bibnamefont
  {Hummer}},\ }\href@noop {} {\bibfield  {journal} {\bibinfo  {journal} {J.
  Struct. Biol.}\ }\textbf {\bibinfo {volume} {184}},\ \bibinfo {pages} {427}
  (\bibinfo {year} {2013})}\BibitemShut {NoStop}%
\bibitem [{\citenamefont {Schneidman-Duhovny}, \citenamefont {Pellarin},\ and\
  \citenamefont {Sali}(2014)}]{Schneidman-Duhovny:14}%
  \BibitemOpen
  \bibfield  {author} {\bibinfo {author} {\bibfnamefont {D.}~\bibnamefont
  {Schneidman-Duhovny}}, \bibinfo {author} {\bibfnamefont {R.}~\bibnamefont
  {Pellarin}}, \ and\ \bibinfo {author} {\bibfnamefont {A.}~\bibnamefont
  {Sali}},\ }\href@noop {} {\bibfield  {journal} {\bibinfo  {journal} {Curr.
  Opin. Struct. Biol.}\ }\textbf {\bibinfo {volume} {28}},\ \bibinfo {pages}
  {96} (\bibinfo {year} {2014})}\BibitemShut {NoStop}%
\bibitem [{\citenamefont {MacKay}(2003)}]{McKayBook}%
  \BibitemOpen
  \bibfield  {author} {\bibinfo {author} {\bibfnamefont {D.~J.~C.}\
  \bibnamefont {MacKay}},\ }\href@noop {} {\emph {\bibinfo {title} {Information
  Theory, Inference, and Learning Algorithms}}}\ (\bibinfo  {publisher}
  {Cambridge University Press},\ \bibinfo {address} {Cambridge, UK},\ \bibinfo
  {year} {2003})\BibitemShut {NoStop}%
\bibitem [{\citenamefont {Rieping}, \citenamefont {Habeck},\ and\ \citenamefont
  {Nilges}(2005)}]{Rieping:05}%
  \BibitemOpen
  \bibfield  {author} {\bibinfo {author} {\bibfnamefont {W.}~\bibnamefont
  {Rieping}}, \bibinfo {author} {\bibfnamefont {M.}~\bibnamefont {Habeck}}, \
  and\ \bibinfo {author} {\bibfnamefont {M.}~\bibnamefont {Nilges}},\
  }\href@noop {} {\bibfield  {journal} {\bibinfo  {journal} {Science}\ }\textbf
  {\bibinfo {volume} {309}},\ \bibinfo {pages} {303} (\bibinfo {year}
  {2005})}\BibitemShut {NoStop}%
\bibitem [{\citenamefont {Crooks}(2007)}]{Crooks:2007}%
  \BibitemOpen
  \bibfield  {author} {\bibinfo {author} {\bibfnamefont {G.~E.}\ \bibnamefont
  {Crooks}},\ }\href {\doibase 10.1103/PhysRevE.75.041119} {\bibfield
  {journal} {\bibinfo  {journal} {Phys. Rev. E}\ }\textbf {\bibinfo {volume}
  {75}},\ \bibinfo {pages} {041119} (\bibinfo {year} {2007})}\BibitemShut
  {NoStop}%
\bibitem [{\citenamefont {Gull}\ and\ \citenamefont {Daniell}(1978)}]{Gull:78}%
  \BibitemOpen
  \bibfield  {author} {\bibinfo {author} {\bibfnamefont {S.~F.}\ \bibnamefont
  {Gull}}\ and\ \bibinfo {author} {\bibfnamefont {G.~J.}\ \bibnamefont
  {Daniell}},\ }\href@noop {} {\bibfield  {journal} {\bibinfo  {journal}
  {Nature}\ }\textbf {\bibinfo {volume} {272}},\ \bibinfo {pages} {686}
  (\bibinfo {year} {1978})}\BibitemShut {NoStop}%
\bibitem [{\citenamefont {Jaynes}(1982)}]{Jaynes:1982}%
  \BibitemOpen
  \bibfield  {author} {\bibinfo {author} {\bibfnamefont {E.~T.}\ \bibnamefont
  {Jaynes}},\ }\href@noop {} {\bibfield  {journal} {\bibinfo  {journal} {Proc.
  IEEE}\ }\textbf {\bibinfo {volume} {70}},\ \bibinfo {pages} {952} (\bibinfo
  {year} {1982})}\BibitemShut {NoStop}%
\bibitem [{\citenamefont {Press}\ \emph {et~al.}(1992)\citenamefont {Press},
  \citenamefont {Teukolsky}, \citenamefont {Vetterling},\ and\ \citenamefont
  {Flannery}}]{MaxEntNumericalRecipes}%
  \BibitemOpen
  \bibfield  {author} {\bibinfo {author} {\bibfnamefont {W.~H.}\ \bibnamefont
  {Press}}, \bibinfo {author} {\bibfnamefont {S.~A.}\ \bibnamefont
  {Teukolsky}}, \bibinfo {author} {\bibfnamefont {W.~T.}\ \bibnamefont
  {Vetterling}}, \ and\ \bibinfo {author} {\bibfnamefont {B.~P.}\ \bibnamefont
  {Flannery}},\ }\enquote {\bibinfo {title} {Numerical recipes in {FORTRAN}},}\
  \ (\bibinfo  {publisher} {Cambridge University Press},\ \bibinfo {address}
  {Cambridge, U.K.},\ \bibinfo {year} {1992})\ Chap.\ \bibinfo {chapter}
  {18.7},\ \bibinfo {edition} {2nd}\ ed.\BibitemShut {Stop}%
\bibitem [{\citenamefont {Pitera}\ and\ \citenamefont
  {Chodera}(2012)}]{Pitera:12}%
  \BibitemOpen
  \bibfield  {author} {\bibinfo {author} {\bibfnamefont {J.~W.}\ \bibnamefont
  {Pitera}}\ and\ \bibinfo {author} {\bibfnamefont {J.~D.}\ \bibnamefont
  {Chodera}},\ }\href@noop {} {\bibfield  {journal} {\bibinfo  {journal} {J.
  Chem. Theory Comput.}\ }\textbf {\bibinfo {volume} {8}},\ \bibinfo {pages}
  {3445} (\bibinfo {year} {2012})}\BibitemShut {NoStop}%
\bibitem [{\citenamefont {White}\ and\ \citenamefont {Voth}(2014)}]{White:14}%
  \BibitemOpen
  \bibfield  {author} {\bibinfo {author} {\bibfnamefont {A.~D.}\ \bibnamefont
  {White}}\ and\ \bibinfo {author} {\bibfnamefont {G.~A.}\ \bibnamefont
  {Voth}},\ }\href@noop {} {\bibfield  {journal} {\bibinfo  {journal} {J. Chem.
  Theory Comput.}\ }\textbf {\bibinfo {volume} {10}},\ \bibinfo {pages} {3023}
  (\bibinfo {year} {2014})}\BibitemShut {NoStop}%
\bibitem [{\citenamefont {White}, \citenamefont {Dama},\ and\ \citenamefont
  {Voth}(2015)}]{White:15}%
  \BibitemOpen
  \bibfield  {author} {\bibinfo {author} {\bibfnamefont {A.~D.}\ \bibnamefont
  {White}}, \bibinfo {author} {\bibfnamefont {J.~F.}\ \bibnamefont {Dama}}, \
  and\ \bibinfo {author} {\bibfnamefont {G.~A.}\ \bibnamefont {Voth}},\
  }\href@noop {} {\bibfield  {journal} {\bibinfo  {journal} {J. Chem. Theory
  Comput.}\ }\textbf {\bibinfo {volume} {11}},\ \bibinfo {pages} {2451}
  (\bibinfo {year} {2015})}\BibitemShut {NoStop}%
\bibitem [{\citenamefont {Beauchamp}, \citenamefont {Pande},\ and\
  \citenamefont {Das}(2014)}]{Beauchamp:14}%
  \BibitemOpen
  \bibfield  {author} {\bibinfo {author} {\bibfnamefont {K.~A.}\ \bibnamefont
  {Beauchamp}}, \bibinfo {author} {\bibfnamefont {V.~S.}\ \bibnamefont
  {Pande}}, \ and\ \bibinfo {author} {\bibfnamefont {R.}~\bibnamefont {Das}},\
  }\href@noop {} {\bibfield  {journal} {\bibinfo  {journal} {Biophys. J.}\
  }\textbf {\bibinfo {volume} {106}},\ \bibinfo {pages} {1381} (\bibinfo {year}
  {2014})}\BibitemShut {NoStop}%
\bibitem [{\citenamefont {Kim}\ and\ \citenamefont
  {Prestegard}(1989)}]{Kim:89}%
  \BibitemOpen
  \bibfield  {author} {\bibinfo {author} {\bibfnamefont {Y.}~\bibnamefont
  {Kim}}\ and\ \bibinfo {author} {\bibfnamefont {J.~H.}\ \bibnamefont
  {Prestegard}},\ }\href@noop {} {\bibfield  {journal} {\bibinfo  {journal}
  {Biochemistry}\ }\textbf {\bibinfo {volume} {28}},\ \bibinfo {pages} {8792}
  (\bibinfo {year} {1989})}\BibitemShut {NoStop}%
\bibitem [{\citenamefont {Kuriyan}\ \emph {et~al.}(1991)\citenamefont
  {Kuriyan}, \citenamefont {Osapay}, \citenamefont {Burley}, \citenamefont
  {Brunger}, \citenamefont {Hendrickson},\ and\ \citenamefont
  {Karplus}}]{Kuriyan:91}%
  \BibitemOpen
  \bibfield  {author} {\bibinfo {author} {\bibfnamefont {J.}~\bibnamefont
  {Kuriyan}}, \bibinfo {author} {\bibfnamefont {K.}~\bibnamefont {Osapay}},
  \bibinfo {author} {\bibfnamefont {S.~K.}\ \bibnamefont {Burley}}, \bibinfo
  {author} {\bibfnamefont {A.~T.}\ \bibnamefont {Brunger}}, \bibinfo {author}
  {\bibfnamefont {W.~A.}\ \bibnamefont {Hendrickson}}, \ and\ \bibinfo {author}
  {\bibfnamefont {M.}~\bibnamefont {Karplus}},\ }\href@noop {} {\bibfield
  {journal} {\bibinfo  {journal} {Proteins Struct. Funct. Genet.}\ }\textbf
  {\bibinfo {volume} {10}},\ \bibinfo {pages} {340} (\bibinfo {year}
  {1991})}\BibitemShut {NoStop}%
\bibitem [{\citenamefont {Best}\ and\ \citenamefont
  {Vendruscolo}(2004)}]{Best:JACS:2004}%
  \BibitemOpen
  \bibfield  {author} {\bibinfo {author} {\bibfnamefont {R.~B.}\ \bibnamefont
  {Best}}\ and\ \bibinfo {author} {\bibfnamefont {M.}~\bibnamefont
  {Vendruscolo}},\ }\href@noop {} {\bibfield  {journal} {\bibinfo  {journal}
  {J. Am. Chem. Soc.}\ }\textbf {\bibinfo {volume} {126}},\ \bibinfo {pages}
  {8090} (\bibinfo {year} {2004})}\BibitemShut {NoStop}%
\bibitem [{\citenamefont {Lindorff-Larsen}\ \emph {et~al.}(2005)\citenamefont
  {Lindorff-Larsen}, \citenamefont {Best}, \citenamefont {Depristo},
  \citenamefont {Dobson},\ and\ \citenamefont
  {Vendruscolo}}]{Lindorff-Larsen:05}%
  \BibitemOpen
  \bibfield  {author} {\bibinfo {author} {\bibfnamefont {K.}~\bibnamefont
  {Lindorff-Larsen}}, \bibinfo {author} {\bibfnamefont {R.~B.}\ \bibnamefont
  {Best}}, \bibinfo {author} {\bibfnamefont {M.~A.}\ \bibnamefont {Depristo}},
  \bibinfo {author} {\bibfnamefont {C.~M.}\ \bibnamefont {Dobson}}, \ and\
  \bibinfo {author} {\bibfnamefont {M.}~\bibnamefont {Vendruscolo}},\
  }\href@noop {} {\bibfield  {journal} {\bibinfo  {journal} {Nature}\ }\textbf
  {\bibinfo {volume} {433}},\ \bibinfo {pages} {128} (\bibinfo {year}
  {2005})}\BibitemShut {NoStop}%
\bibitem [{\citenamefont {Cavalli}, \citenamefont {Camilloni},\ and\
  \citenamefont {Vendruscolo}(2013{\natexlab{a}})}]{Cavalli:13:1}%
  \BibitemOpen
  \bibfield  {author} {\bibinfo {author} {\bibfnamefont {A.}~\bibnamefont
  {Cavalli}}, \bibinfo {author} {\bibfnamefont {C.}~\bibnamefont {Camilloni}},
  \ and\ \bibinfo {author} {\bibfnamefont {M.}~\bibnamefont {Vendruscolo}},\
  }\href@noop {} {\bibfield  {journal} {\bibinfo  {journal} {J. Chem. Phys.}\
  }\textbf {\bibinfo {volume} {138}},\ \bibinfo {pages} {094112} (\bibinfo
  {year} {2013}{\natexlab{a}})}\BibitemShut {NoStop}%
\bibitem [{\citenamefont {Cavalli}, \citenamefont {Camilloni},\ and\
  \citenamefont {Vendruscolo}(2013{\natexlab{b}})}]{Cavalli:Erratum:13}%
  \BibitemOpen
  \bibfield  {author} {\bibinfo {author} {\bibfnamefont {A.}~\bibnamefont
  {Cavalli}}, \bibinfo {author} {\bibfnamefont {C.}~\bibnamefont {Camilloni}},
  \ and\ \bibinfo {author} {\bibfnamefont {M.}~\bibnamefont {Vendruscolo}},\
  }\href@noop {} {\bibfield  {journal} {\bibinfo  {journal} {J. Chem. Phys.}\
  }\textbf {\bibinfo {volume} {139}},\ \bibinfo {pages} {169903} (\bibinfo
  {year} {2013}{\natexlab{b}})}\BibitemShut {NoStop}%
\bibitem [{\citenamefont {Bonomi}\ \emph {et~al.}(2015)\citenamefont {Bonomi},
  \citenamefont {Camilloni}, \citenamefont {Cavalli},\ and\ \citenamefont
  {Vendruscolo}}]{Bonomi:15}%
  \BibitemOpen
  \bibfield  {author} {\bibinfo {author} {\bibfnamefont {M.}~\bibnamefont
  {Bonomi}}, \bibinfo {author} {\bibfnamefont {C.}~\bibnamefont {Camilloni}},
  \bibinfo {author} {\bibfnamefont {A.}~\bibnamefont {Cavalli}}, \ and\
  \bibinfo {author} {\bibfnamefont {M.}~\bibnamefont {Vendruscolo}},\
  }\href@noop {} {\enquote {\bibinfo {title} {Metainference: {A} {Bayesian}
  inference method for heterogeneous systems},}\ }\bibinfo {howpublished}
  {http://arxiv.org/abs/1509.05684} (\bibinfo {year} {2015})\BibitemShut
  {NoStop}%
\bibitem [{\citenamefont {Roux}\ and\ \citenamefont
  {Weare}(2013)}]{RouxWeare:13}%
  \BibitemOpen
  \bibfield  {author} {\bibinfo {author} {\bibfnamefont {B.}~\bibnamefont
  {Roux}}\ and\ \bibinfo {author} {\bibfnamefont {J.}~\bibnamefont {Weare}},\
  }\href@noop {} {\bibfield  {journal} {\bibinfo  {journal} {J. Chem. Phys.}\
  }\textbf {\bibinfo {volume} {138}},\ \bibinfo {pages} {084107} (\bibinfo
  {year} {2013})}\BibitemShut {NoStop}%
\bibitem [{\citenamefont {Rosta}\ \emph {et~al.}(2011)\citenamefont {Rosta},
  \citenamefont {Nowotny}, \citenamefont {Yang},\ and\ \citenamefont
  {Hummer}}]{Rosta:JACS:2011}%
  \BibitemOpen
  \bibfield  {author} {\bibinfo {author} {\bibfnamefont {E.}~\bibnamefont
  {Rosta}}, \bibinfo {author} {\bibfnamefont {M.}~\bibnamefont {Nowotny}},
  \bibinfo {author} {\bibfnamefont {W.}~\bibnamefont {Yang}}, \ and\ \bibinfo
  {author} {\bibfnamefont {G.}~\bibnamefont {Hummer}},\ }\href@noop {}
  {\bibfield  {journal} {\bibinfo  {journal} {J. Am. Chem. Soc.}\ }\textbf
  {\bibinfo {volume} {133}},\ \bibinfo {pages} {8934} (\bibinfo {year}
  {2011})}\BibitemShut {NoStop}%
\bibitem [{\citenamefont {Shirts}\ and\ \citenamefont
  {Chodera}(2008)}]{Shirts:08}%
  \BibitemOpen
  \bibfield  {author} {\bibinfo {author} {\bibfnamefont {M.~R.}\ \bibnamefont
  {Shirts}}\ and\ \bibinfo {author} {\bibfnamefont {J.~D.}\ \bibnamefont
  {Chodera}},\ }\href@noop {} {\bibfield  {journal} {\bibinfo  {journal} {J.
  Chem. Phys.}\ }\textbf {\bibinfo {volume} {129}},\ \bibinfo {pages} {124105}
  (\bibinfo {year} {2008})}\BibitemShut {NoStop}%
\bibitem [{\citenamefont {Souaille}\ and\ \citenamefont
  {Roux}(2001)}]{Souaille:01}%
  \BibitemOpen
  \bibfield  {author} {\bibinfo {author} {\bibfnamefont {M.}~\bibnamefont
  {Souaille}}\ and\ \bibinfo {author} {\bibfnamefont {B.}~\bibnamefont
  {Roux}},\ }\href@noop {} {\bibfield  {journal} {\bibinfo  {journal} {Comp.
  Phys. Comm.}\ }\textbf {\bibinfo {volume} {135}},\ \bibinfo {pages} {40}
  (\bibinfo {year} {2001})}\BibitemShut {NoStop}%
\bibitem [{\citenamefont {Peter}, \citenamefont {Daura},\ and\ \citenamefont
  {{van Gunsteren}}(2001)}]{Peter:01}%
  \BibitemOpen
  \bibfield  {author} {\bibinfo {author} {\bibfnamefont {C.}~\bibnamefont
  {Peter}}, \bibinfo {author} {\bibfnamefont {X.}~\bibnamefont {Daura}}, \ and\
  \bibinfo {author} {\bibfnamefont {W.~F.}\ \bibnamefont {{van Gunsteren}}},\
  }\href@noop {} {\bibfield  {journal} {\bibinfo  {journal} {J. Biomol. NMR}\
  }\textbf {\bibinfo {volume} {20}},\ \bibinfo {pages} {297} (\bibinfo {year}
  {2001})}\BibitemShut {NoStop}%
\bibitem [{\citenamefont {Hansen}\ and\ \citenamefont
  {{O'Leary}}(1993)}]{Hansen:93}%
  \BibitemOpen
  \bibfield  {author} {\bibinfo {author} {\bibfnamefont {P.~C.}\ \bibnamefont
  {Hansen}}\ and\ \bibinfo {author} {\bibfnamefont {D.~P.}\ \bibnamefont
  {{O'Leary}}},\ }\href@noop {} {\bibfield  {journal} {\bibinfo  {journal}
  {SIAM J. Sci. Comput.}\ }\textbf {\bibinfo {volume} {14}},\ \bibinfo {pages}
  {1487} (\bibinfo {year} {1993})}\BibitemShut {NoStop}%
\bibitem [{\citenamefont {Camilloni}, \citenamefont {Cavalli},\ and\
  \citenamefont {Vendruscolo}(2013)}]{Camilloni:JCTC:13}%
  \BibitemOpen
  \bibfield  {author} {\bibinfo {author} {\bibfnamefont {C.}~\bibnamefont
  {Camilloni}}, \bibinfo {author} {\bibfnamefont {A.}~\bibnamefont {Cavalli}},
  \ and\ \bibinfo {author} {\bibfnamefont {M.}~\bibnamefont {Vendruscolo}},\
  }\href@noop {} {\bibfield  {journal} {\bibinfo  {journal} {J. Chem. Theory
  Comput.}\ }\textbf {\bibinfo {volume} {9}},\ \bibinfo {pages} {5610}
  (\bibinfo {year} {2013})}\BibitemShut {NoStop}%
\bibitem [{\citenamefont {Hansen}\ \emph {et~al.}(2014)\citenamefont {Hansen},
  \citenamefont {Heller}, \citenamefont {Schmid},\ and\ \citenamefont {{van
  Gunsteren}}}]{Hansen:14}%
  \BibitemOpen
  \bibfield  {author} {\bibinfo {author} {\bibfnamefont {N.}~\bibnamefont
  {Hansen}}, \bibinfo {author} {\bibfnamefont {F.}~\bibnamefont {Heller}},
  \bibinfo {author} {\bibfnamefont {N.}~\bibnamefont {Schmid}}, \ and\ \bibinfo
  {author} {\bibfnamefont {W.~F.}\ \bibnamefont {{van Gunsteren}}},\
  }\href@noop {} {\bibfield  {journal} {\bibinfo  {journal} {J. Biomol. NMR}\
  }\textbf {\bibinfo {volume} {60}},\ \bibinfo {pages} {169} (\bibinfo {year}
  {2014})}\BibitemShut {NoStop}%
\bibitem [{\citenamefont {Marinelli}\ and\ \citenamefont
  {Faraldo-Gomez}(2015)}]{Marinelli:15}%
  \BibitemOpen
  \bibfield  {author} {\bibinfo {author} {\bibfnamefont {F.}~\bibnamefont
  {Marinelli}}\ and\ \bibinfo {author} {\bibfnamefont {J.~D.}\ \bibnamefont
  {Faraldo-Gomez}},\ }\href@noop {} {\bibfield  {journal} {\bibinfo  {journal}
  {Biophys. J.}\ }\textbf {\bibinfo {volume} {108}},\ \bibinfo {pages} {2779}
  (\bibinfo {year} {2015})}\BibitemShut {NoStop}%
\bibitem [{\citenamefont {Norgaard}, \citenamefont {Ferkinghoff-Borg},\ and\
  \citenamefont {Lindorff-Larsen}(2008)}]{Norgaard:08}%
  \BibitemOpen
  \bibfield  {author} {\bibinfo {author} {\bibfnamefont {A.~B.}\ \bibnamefont
  {Norgaard}}, \bibinfo {author} {\bibfnamefont {J.}~\bibnamefont
  {Ferkinghoff-Borg}}, \ and\ \bibinfo {author} {\bibfnamefont
  {K.}~\bibnamefont {Lindorff-Larsen}},\ }\href@noop {} {\bibfield  {journal}
  {\bibinfo  {journal} {Biophys. J.}\ }\textbf {\bibinfo {volume} {94}},\
  \bibinfo {pages} {182} (\bibinfo {year} {2008})}\BibitemShut {NoStop}%
\bibitem [{\citenamefont {Li}\ and\ \citenamefont
  {Brueschweiler}(2011)}]{Li:11}%
  \BibitemOpen
  \bibfield  {author} {\bibinfo {author} {\bibfnamefont {D.~W.}\ \bibnamefont
  {Li}}\ and\ \bibinfo {author} {\bibfnamefont {R.}~\bibnamefont
  {Brueschweiler}},\ }\href@noop {} {\bibfield  {journal} {\bibinfo  {journal}
  {J. Chem. Theory Comput.}\ }\textbf {\bibinfo {volume} {7}},\ \bibinfo
  {pages} {1773} (\bibinfo {year} {2011})}\BibitemShut {NoStop}%
\bibitem [{\citenamefont {Wang}, \citenamefont {Chen},\ and\ \citenamefont
  {Voorhis}(2013)}]{Wang:13}%
  \BibitemOpen
  \bibfield  {author} {\bibinfo {author} {\bibfnamefont {L.-P.}\ \bibnamefont
  {Wang}}, \bibinfo {author} {\bibfnamefont {J.}~\bibnamefont {Chen}}, \ and\
  \bibinfo {author} {\bibfnamefont {T.}\ \bibnamefont {Van Voorhis}},\
  }\href@noop {} {\bibfield  {journal} {\bibinfo  {journal} {J. Chem. Theory
  Comput.}\ }\textbf {\bibinfo {volume} {9}},\ \bibinfo {pages} {452}
  (\bibinfo {year} {2013})}\BibitemShut {NoStop}%
\bibitem [{\citenamefont {Best}\ and\ \citenamefont {Hummer}(2009)}]{Best:09}%
  \BibitemOpen
  \bibfield  {author} {\bibinfo {author} {\bibfnamefont {R.~B.}\ \bibnamefont
  {Best}}\ and\ \bibinfo {author} {\bibfnamefont {G.}~\bibnamefont {Hummer}},\
  }\href@noop {} {\bibfield  {journal} {\bibinfo  {journal} {J. Phys. Chem. B}\
  }\textbf {\bibinfo {volume} {113}},\ \bibinfo {pages} {9004} (\bibinfo {year}
  {2009})}\BibitemShut {NoStop}%
\bibitem [{\citenamefont {Chen}\ and\ \citenamefont
  {Garc{\'{\i}}a}(2013)}]{Chen:13}%
  \BibitemOpen
  \bibfield  {author} {\bibinfo {author} {\bibfnamefont {A.~A.}\ \bibnamefont
  {Chen}}\ and\ \bibinfo {author} {\bibfnamefont {A.~E.}\ \bibnamefont
  {Garc{\'{\i}}a}},\ }\href@noop {} {\bibfield  {journal} {\bibinfo  {journal}
  {Proc. Natl. Acad. Sci. U.S.A.}\ }\textbf {\bibinfo {volume} {110}},\
  \bibinfo {pages} {16820} (\bibinfo {year} {2013})}\BibitemShut {NoStop}%
\bibitem [{\citenamefont {Habeck}(2014)}]{Habeck:14}%
  \BibitemOpen
  \bibfield  {author} {\bibinfo {author} {\bibfnamefont {M.}~\bibnamefont
  {Habeck}},\ }\href@noop {} {\bibfield  {journal} {\bibinfo  {journal} {Phys.
  Rev. E}\ }\textbf {\bibinfo {volume} {89}},\ \bibinfo {pages} {052113}
  (\bibinfo {year} {2014})}\BibitemShut {NoStop}%
\bibitem [{\citenamefont {Best}, \citenamefont {Chen},\ and\ \citenamefont
  {Hummer}(2005)}]{best-2005-4}%
  \BibitemOpen
  \bibfield  {author} {\bibinfo {author} {\bibfnamefont {R.~B.}\ \bibnamefont
  {Best}}, \bibinfo {author} {\bibfnamefont {Y.-G.}\ \bibnamefont {Chen}}, \
  and\ \bibinfo {author} {\bibfnamefont {G.}~\bibnamefont {Hummer}},\
  }\href@noop {} {\bibfield  {journal} {\bibinfo  {journal} {Structure}\
  }\textbf {\bibinfo {volume} {13}},\ \bibinfo {pages} {1755} (\bibinfo {year}
  {2005})}\BibitemShut {NoStop}%
\bibitem [{\citenamefont {Berlin}\ \emph {et~al.}(2013)\citenamefont {Berlin},
  \citenamefont {Castaneda}, \citenamefont {Schneidman-Duhovny}, \citenamefont
  {Sali}, \citenamefont {Nava-Tudela},\ and\ \citenamefont
  {Fushman}}]{Berlin:13}%
  \BibitemOpen
  \bibfield  {author} {\bibinfo {author} {\bibfnamefont {K.}~\bibnamefont
  {Berlin}}, \bibinfo {author} {\bibfnamefont {C.~A.}\ \bibnamefont
  {Castaneda}}, \bibinfo {author} {\bibfnamefont {D.}~\bibnamefont
  {Schneidman-Duhovny}}, \bibinfo {author} {\bibfnamefont {A.}~\bibnamefont
  {Sali}}, \bibinfo {author} {\bibfnamefont {A.}~\bibnamefont {Nava-Tudela}}, \
  and\ \bibinfo {author} {\bibfnamefont {D.}~\bibnamefont {Fushman}},\
  }\href@noop {} {\bibfield  {journal} {\bibinfo  {journal} {J. Am. Chem.
  Soc.}\ }\textbf {\bibinfo {volume} {135}},\ \bibinfo {pages} {16595}
  (\bibinfo {year} {2013})}\BibitemShut {NoStop}%
\bibitem [{\citenamefont {Pelikan}, \citenamefont {Hura},\ and\ \citenamefont
  {Hammel}(2009)}]{Pelikan:09}%
  \BibitemOpen
  \bibfield  {author} {\bibinfo {author} {\bibfnamefont {M.}~\bibnamefont
  {Pelikan}}, \bibinfo {author} {\bibfnamefont {G.~L.}\ \bibnamefont {Hura}}, \
  and\ \bibinfo {author} {\bibfnamefont {M.}~\bibnamefont {Hammel}},\
  }\href@noop {} {\bibfield  {journal} {\bibinfo  {journal} {Gen. Phys.
  Biophys.}\ }\textbf {\bibinfo {volume} {28}},\ \bibinfo {pages} {174}
  (\bibinfo {year} {2009})}\BibitemShut {NoStop}%
\end{thebibliography}

%

\end{document}